\documentclass[iop,apj,tighten]{emulateapj}
\usepackage{epsfig}
\usepackage{amsmath}
\usepackage{natbib}
\usepackage{url}
\usepackage{hyperref}
\usepackage{array}

\usepackage{datetime} 
\usepackage{color}
\definecolor{gris}{gray}{0.65}
\definecolor{blue}{rgb}{0.2,0,0.6}
\definecolor{violet}{rgb}{0.5,0,0.5}
\definecolor{aquadark}{rgb}{0,0.5,0.5}
\definecolor{aquapale}{rgb}{0,0.7,0.8}
\definecolor{darkpink}{rgb}{0.8,0.2,0.4}
\definecolor{pink}{rgb}{1,0.2,0.2}
\definecolor{palegreen}{rgb}{0,0.5,0}
\definecolor{yellow}{rgb}{0.9,0.9,0.5}
\definecolor{red}{rgb}{0.9,0,0}
\definecolor{orange}{rgb}{0.9,0.6,0.2} 

\usepackage{soul} 
\usepackage{ulem}
\newcommand{\remove}[1]{}

\def \Msun {$M_\odot$}
\def \MJ {$\,M_{\rm{Jup}}$}
\def \RJ {$\,R_{\rm{Jup}}$}
\def \Teff {$T_{\rm{eff}}$}
\def \logg {$\log g$}
\def \logLx {$\log$\,$L_{X}$}
\def \fsed {$f_{\rm{sed}}$}
\def \Kzz {$K_{zz}$}
\def \Lbol {$\log (L/L_{\odot})$}
\def \ds {$d_{s}$}
\def \vsini {$v\sin i$}
\def \Ebind {$E_{\rm{bind}}$}
\def \vrad {$v_{\rm{rad}}$}
\def \Prot {$P_{\rm{rot}}$}
\def \EWHa {EW$_{\rm{H}{\alpha}}$}
\def \Ha {H$\alpha$}
\def \CHq {$\rm{CH}_{4}$}
\def \Hd {$\rm{H}_{2}$}
\def \HdO {$\rm{H}_{2}O$}
\def \NHt {$\rm{NH}_{3}$}
\def \env {$\sim$\,}
\def \plmo {\,$\pm$\,}
\def \gta {\,\lower 0.5ex\hbox{$\buildrel > \over \sim\ $}\,}   
\def \lta {\,\lower 0.5ex\hbox{$\buildrel < \over \sim\ $}\,}   
\def \deg {$^{\circ}$}
\def \smun {s$^{-1}$}
\def \kms {km s$^{-1}$}
\def \cmdsmun {cm$^{2}$s$^{-1}$}
\def \Ks {$K_{\rm s}$}
\def \BPMG {$\beta$PMG}
\usepackage[mediumspace,mediumqspace,Grey,squaren]{SIunits}
\def \micron {\micro\meter}
\def \tnma {\tablenotemark{a}}
\def \tnmb {\tablenotemark{b}}
\def \tnmc {\tablenotemark{c}}
\def \tnmd {\tablenotemark{d}}
\def \tnme {\tablenotemark{e}}
\def \tnmf {\tablenotemark{f}}
\def \tnmg {\tablenotemark{g}}
\def \tnmh {\tablenotemark{h}}
\def \tnmi {\tablenotemark{i}}
\def \tnmj {\tablenotemark{j}}
\def \tnmk {\tablenotemark{k}}
\def \tnml {\tablenotemark{l}}
\def \tnmm {\tablenotemark{m}}
\def \tnmn {\tablenotemark{n}}
\def \tnmo {\tablenotemark{o}}
\def \tnmp {\tablenotemark{p}}
\def \tnmq {\tablenotemark{q}}
\def \tnmr {\tablenotemark{r}}
\def \tnms {\tablenotemark{s}}

\def \Maloprep {\citet{Malo2014arxiv}}
\def \Maloprepalt {\citealt{Malo2014arxiv}}
\def \Naudprep {(M.-E. Naud et al., in preparation)}
\def \Artigauprep {(\'{E}. Artigau et al., in preparation)}
\def \Gagne {\citet{Gagne2013arxiv2}}

\begin{document}
\title{DISCOVERY OF A WIDE PLANETARY-MASS COMPANION TO THE YOUNG M3 STAR GU Psc}
\author{Marie-Eve Naud$^1$*, \'{E}tienne Artigau$^1$, Lison Malo$^1$, Lo\"{i}c Albert$^1$, Ren\'{e} Doyon$^1$, David Lafreni\`{e}re$^1$, Jonathan Gagn\'{e}$^1$, Didier Saumon$^2$, Caroline V. Morley$^3$, France Allard$^4$, Derek Homeier$^{4}$, Charles A. Beichman$^5$, Christopher R. Gelino$^{5,6}$, and Anne Boucher$^1$}
\email{*Corresponding author: naud@astro.umontreal.ca}
\affil{$^1$D\'{e}partement de physique and Observatoire du Mont-M\'{e}gantic, Universit\'{e} de Montr\'{e}al, Montr\'{e}al H3C 3J7, Canada}
\affil{$^2$Los Alamos National Laboratory, Los Alamos, NM 87545, USA}
\affil{$^3$Department of Astronomy and Astrophysics, University of California, Santa Cruz, CA 95064, USA}
\affil{$^4$Centre de Recherche Astrophysique de Lyon, UMR 5574 CNRS, Universit\'{e} de Lyon, \'{E}cole Normale Sup\'{e}rieure de Lyon, 46 All\'{e}e d'Italie, F-69364 Lyon Cedex 07, France}
\affil{$^5$ Infrared Processing and Analysis Center, MS 100-22, California Institute of Technology, Pasadena, CA 91125, USA}
\affil{$^6$ NASA Exoplanet Science Institute, California Institute of Technology, 770 S. Wilson Ave., Pasadena, CA 91125, USA}

\begin{abstract}
We present the discovery of a co-moving planetary-mass companion \env42\arcsec\ (\env2000\,AU) from a young M3 star, GU Psc, likely member of the young AB Doradus Moving Group (ABDMG). The companion was first identified via its distinctively red $i$-$z$ color ($>$ 3.5) through a survey made with Gemini-S/GMOS. Follow-up Canada--France--Hawaii Telescope/WIRCam near-infrared (NIR) imaging, Gemini-N/GNIRS NIR spectroscopy and \textit{Wide-field Infrared Survey Explorer} photometry indicate a spectral type of T3.5\plmo1 and reveal signs of low gravity which we attribute to youth. Keck/Adaptive Optics NIR observations did not resolve the companion as a binary. A comparison with atmosphere models indicates \Teff\ = 1000--1100\,K and \logg\ = 4.5--5.0. Based on evolution models, this temperature corresponds to a mass of 9--13\,\MJ for the age of ABDMG (70--130\,Myr). The relatively well-constrained age of this companion and its very large angular separation to its host star will allow its thorough characterization and will make it a valuable comparison for planetary-mass companions that will be uncovered by forthcoming planet-finder instruments such as Gemini Planet Imager and SPHERE. 
\end{abstract}
\keywords{planetary systemsÐplanets and satellites: detectionÐstars: individual (GU Psc)Ð stars: low-massÐstars: imagingÐinfrared: planetary systems}

\section{Introduction}
Of the nearly thousand of exoplanets known so far, the majority ($>$ 90\%) were detected through the radial velocity and the transit methods\footnote{The Extrasolar Planets Encyclopaedia \citep{Schneider2011}, \url{exoplanet.eu}.}. The sample is thus biased toward planets at relatively small orbital separations of a few astronomical units or less. Direct imaging complements these methods by finding the most massive planets at large orbital separations. Most of the directly imaged planets known today (e.g., HR8799bcde, \citealt{Marois2008,Marois2010}; $\beta$ Pictoris b, \citealt{Lagrange2009}; 1RXS J1609-2105b, \citealt{Lafreniere2008, Lafreniere2010}; GJ 504 b, \citealt{Kuzuhara2013}; HD 95086 b, \citealt{Rameau2013}) were found using high-contrast imaging techniques and Adaptive Optics (AO). 

The recent discoveries of low-mass companions at very large orbital separations through seeing-limited imaging came somewhat as a surprise and provided new constraints to formation models. For example, Ross 458(AB) c is a late T dwarf located 1100\,AU (102\arcsec) from its parent pair of M dwarfs \citep{Goldman2010}. Its estimated mass is below the deuterium-burning limit (\env13\,\MJ; \citealt{Spiegel2011a}), a criteria commonly used as the delineation between planets and brown dwarfs. HN Peg b is a more massive (22\plmo9\,\MJ) T2 companion located 795\,AU (43\farcs2) from a G dwarf of 0.3\plmo0.2\,Gyr \citep{Luhman2007}. The planetary-mass companion to the young (8--20\,Myr) brown dwarf 2MASS J12073346-3932539 (2M1207 hereafter; \citealt{Gizis2002,Chauvin2004,Ducourant2008}) could also be added as an example. For a more complete list, see \citet{Neuhauser2012}. 

The wide orbital distances of these companions preclude in situ formation in a protoplanetary disk, which is normally expected for planets. They could have been ejected at this distance through dynamical interactions, or formed like brown dwarfs and stars through collapse and fragmentation of a molecular cloud core. These distant objects are not only easier to detect but also easier to study spectroscopically. They thus constitute excellent proxies to improve atmospheric models and better understand closer-in companions found with AO. For example, the study of Ross 458 (AB) c \citep{Burningham2011,Burgasser2010} suggested that condensate opacity plays a role in the spectra of late T dwarfs and showed that including them allows a better determination of the physical parameters of these objects. Such a detailed spectroscopic characterization is very challenging for planetary-mass companion found very close to their parent star (e.g., HR8799b and c; \citealt{Barman2011,Konopacky2013}).

Young stars are really interesting for direct imaging because their planets, still contracting, are hotter and more luminous than their older counterpart and thus are easier to detect. In the solar neighborhood, young stars are often found in young moving groups and associations \citep{ZuckermanSong2004}. These groups of stars share a common origin and thus have similar positions and space motions in the Galaxy. Since the determination of the mass of directly imaged companions relies on the use of evolutionary models, these young association members, with their constrained age, are prime targets for direct imaging investigation. The lower-mass members of those groups are great targets for imaging, their faint luminosity results in a higher contrast for a planet of a given mass. 

In 2011, we initiated a survey with Gemini-South/GMOS \citep{Hook2004} to search for low-mass companions via their distinctively red $i$-$z$ colors around candidate members of nearby young ($<$ 150\,Myr) associations \Artigauprep. The 91 targets of this survey are low-mass stars (K5--M5) that were recently identified as likely members of nearby young moving groups by \citet{Malo2013} through a novel Bayesian analysis. This survey allowed to search for companion \gta8\,\MJ\ at separations ranging from \env300 to 5000\,AU. One single candidate companion was identified, around GU Pisces (hereafter GU Psc), an M3\plmo0.5 star candidate member of the young AB Doradus Moving Group (ABDMG; \citealt{Zuckerman2004, Zuckerman2011}).

In this paper, we present new observations of both the host star, GU Psc, and its newly detected companion, showing that they form a bound system, with an age consistent with that of the ABDMG, which suggests, according to evolutionary models, a companion mass below the deuterium-burning limit. In Section \ref{sec:obsred}, we present various observations that were carried on the host star, and then on the companion, and explain the reduction of the associated data. In Section \ref{sec:res}, the physical properties of the host star and of the companion are derived. Finally, in Section \ref{sec:analdis}, the stability of this wide pair and plausible formation scenarios are briefly discussed, and the interest of the companion as a proxy for other, closer-in companions is presented.

\section{Observation and Data Reduction}
\label{sec:obsred}
All astrometric and photometric measurements for both the host star and the companion are reported in Table \ref{table:prop}. We present near-infrared (NIR) photometry in the ÊMauna Kea Observatory system (MKO; \citealt{SimonsTokunaga2002, TokunagaVacca2005}) unless stated otherwise.

\begin{table}[htbp]
 \newcolumntype{A}[1]{>{\arraybackslash}m{#1}} 
 \newcolumntype{W}[1]{>{\centering\arraybackslash}m{#1}}  
\caption{Properties of the GU Psc System}
\label{table:prop}
\begin{center}
\begin{tabular}{W{2.1cm}W{2.9cm}W{3.2cm}} 
\hline \hline      
\rule{0pt}{3ex}Property    & GU Psc       & GU Psc b\\
\hline
\rule{0pt}{4ex}Spectral type & M3\plmo0.5\tnma&T3.5\plmo1\\
Age&\multicolumn{2}{c}{100\plmo30\,Myr\tnmb}\\ 
Distance&\multicolumn{2}{c}{48\plmo5\,pc\tnmc}\\ 
Ang. sep.&\multicolumn{2}{c}{41.97\plmo0$\farcs$03\,\tnmd}\\
Proj. phys. sep. &\multicolumn{2}{c}{2000\plmo200\,AU}\\
\hline
&&\\
R.A. (J2000) & 01 12 35.04 & 01 12 36.48\tnmd\\
 (h m s) &  & \\
Decl. (J2000)  & +17 03 55.7 & +17 04 31.8\tnmd\\ 
(d m s) & & \\ 
$\mu_{\alpha}\cos\delta$, $\mu_{\delta}$ (mas yr$^{-1}$) & 90\,$\pm$\,6, --102\,$\pm$\,6\tnme\ & 98\plmo15, --92\plmo15\tnmf\\
\hline
&&\\
$B$ (mag)&15.30\tnmg&\\
$V$ (mag)&13.55\tnmh&\\
$R$ (mag)&12.92\tnmg&\\
$u'$ (mag)&17.347\plmo0.011\tnmi&\\
$g'$ (mag)&15.499\plmo0.005\tnmi&\\
$r'$ (mag)&13.650\plmo0.003\tnmi&\\
$i'$ (mag)&12.408\plmo0.001\tnmi&\\
$z'$ (mag)&12.786\plmo0.014\tnmi&\\
$I_{C}$ (mag)&11.65\plmo 0.13\tnmj&\\
$i_{\rm{AB}}$ (mag)&&$>$ 25.28 ($3\sigma$)\tnmk\\
$z_{\rm{AB}}$ (mag)&&21.75\plmo0.07\tnmd\\
$Y$ (mag)&&19.4\plmo0.05\tnml\\
$J$ (mag)  &10.211\plmo0.022\tnmm&18.12\plmo0.03\tnmn\\
&&18.15\plmo0.04 (2011/10)\tnml \\
&&18.11\plmo0.03 (2011/12)\tnml \\
&&18.10\plmo0.03 (2012/09)\tnml \\
$H$ (mag)&9.598\plmo0.022\tnmm&17.70\plmo0.03\tnml\\
\Ks\ (mag)&9.345\plmo0.015\tnmm&17.40\plmo0.03\tnml\\
$W1$ (mag)&9.241\plmo0.022\tnmo&17.17\plmo0.33\tnmp\\
$W2$ (mag)&9.130\plmo0.020\tnmo&15.41\plmo0.22\tnmp\\
$W3$ (mag)&9.007\plmo0.030\tnmo &$>$ 12.396\tnmq \\
$W4$ (mag)&$>$ 8.659\tnmq & $>$ 8.505\tnmq\\
$HR1\tnmr$&--0.17\plmo0.24&\\
X-ray counts\tnmr&(6.6\plmo1.7) $\times$ $10^{-2}$&\\
(counts sec$^{-1}$)&&\\
\hline
&&\\
\multicolumn{3}{A{8.3cm}} {\textbf{Notes}}\\ 
\end{tabular}
\begin{tabular}{p{0.1cm}p{8.1cm}}
\tnma & Determined with the TiO5 index \citep{Riaz20062}.\\
\tnmb & As a candidate member of ABDMG (see  Section \,\ref{sec:GUPsc}).\\
\tnmc & Statistical distance in ABDMG from Bayesian analysis (see Section \,\ref{sec:GUPsc}).\\ 
\tnmd & Measured on GMOS $z$-band image.\\ 
\tnme & Average of PPMXL \citep{Roeser2010},  PPMX \citep{Roeser2008}, PM2000 \citep{Ducourant2006}, SDSS 9 \citep{Ahn2012} and SUPERBLINK \citep{Schlieder2012}.\\  
\tnmf & Measured on WIRCam $J$-band images (Oct. 2011 \& Sept. 2012).\\
\tnmg & From the USNO Catalog \citep{Monet2003}.\\
\tnmh & From SuperWASP \citep{Norton2007}.\\
\tnmi & From SDSS Photometric Catalog, DR9 \citep{Ahn2012}. Corresponds roughly to AB mag, except for $u'$ = $u_{\rm{AB}}$+0.04 and $z'$ = $z_{\rm{AB}}$-0.02, according to \url{www.sdss.org/DR7/algorithms/fluxcal.html}.\\
\tnmj & From \Maloprepalt, considering $i'$ from UCAC 4 catalog (APASS).\\
\tnmk & From the non detection in the 2$^{nd}$ epoch GMOS $i$-band image.\\
\tnml & Measured in WIRCam observations.\\ 
\tnmm & From 2MASS \citep{Cutri2003}.\\ 
\tnmn & Average of the 3 epochs WIRCam observations.\\ 
\tnmp & \textit{WISE} All-Sky data Release \citep{Cutri2012}.\\   
\tnmq & Computed from \textit{WISE} images (see  Section \ref{subsec:obsredWISE}).\\  
\tnmr & \textit{WISE} 95\% confidence upper limit.\\  
\tnms & From ROSAT All-Sky Bright Source Catalogue \citep{Voges1999}. \\
\end{tabular}
\end{center}
\end{table}

\subsection{The Host Star, GU Psc}
\subsubsection{High-resolution Spectroscopy}
\label{sec:spectroprim}
High-resolution optical spectroscopy was obtained with ESPaDOnS \citep{Donati2006} at the Canada--France--Hawaii Telescope (CFHT). The data were reduced by the CFHT queue service observing team using the pipeline UPENA 1.0, that uses the Libre-ESpRIT software \citep{Donati1997}. The resulting spectrum goes from 0.37\,\micron\ to 1.05\,\micron\ (40 grating orders) with R \env\ 68,000. 

High-resolution spectroscopy was also acquired for GU Psc  with two different instruments in the NIR with the specific goal of measuring its precise radial velocity. With  CRIRES on the Very Large Telescope (VLT; \citealt{Kaeufl2004}), the 0\farcs4-wide slit was used in an order centered on 1.555\,\micron\ for a resulting $R$ \env\ 50,000. With PHOENIX on Gemini-South \citep{Hinkle2003}, we used the 0\farcs34 slit with the 1.547--1.568\,\micron\ blocking filter and obtained a resolving power of $R$ \env\ 52,000. The NIR spectroscopic data were reduced using standard procedures with a custom IDL pipeline.

\subsubsection{Near-infrared Medium-resolution Spectroscopy}
A 0.8 \textendash\ 2.4\,\micron\ NIR spectrum of the primary was also obtained with SpeX, the medium-resolution spectrograph and imager at NASA InfraRed Telescope Facility \citep{Rayner2003}, using the cross-dispersing mode with the 0\farcs8 slit, under good seeing conditions (\env1\arcsec). The reduction was done using SpeXtool \citep{Cushing2004, Vacca2003},  and  telluric absorption was corrected with the A3 spectroscopic standard star HIP 5310. It was then flux-calibrated by adjusting the $J$ -- $H$ and $J$ -- \Ks\ synthetic colors with the Two Micron All Sky Survey (2MASS) photometry. 

\subsubsection{High-contrast Imaging}
GU Psc was observed with NICI, the high-contrast imager on Gemini-South \citep{Ftaclas2003,Chun2008}, as part of a survey to find closer-in companion \Naudprep. It was observed twice (2011 October 21 and 2012 September 1) in two narrowband filters around 1.6\,\micron\ (\CHq H 4\% S centered at 1.578\,\micron\ and the \CHq H 4\% L at 1.652\,\micron). Each observing sequence is composed of 35 exposures of 3 coadd $\times$ 20.14\,s, taken with the 0\farcs32 focal plane mask. The reduction was carried using the method detailed in \citet{Artigau2008}. 

\subsection{The Companion, GU Psc b}
\begin{figure}[htbp]
\begin{center}
\includegraphics[width=8.5cm]{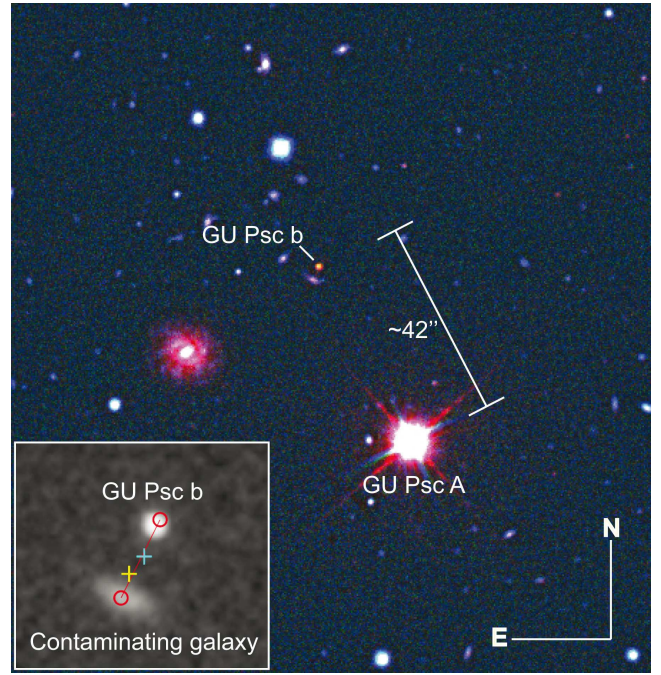}
\caption{Main: composite Gemini-South/GMOS $i$ (blue), $z$ (green), and CFHT/WIRCam $J$ (red) image of GU Psc and its companion. As expected for a substellar object, GU Psc b is much redder in these bands than field stars and most background galaxies. Inset: close-up on GU Psc b and the galaxy located \env3\arcsec\ south east from it. The two red circles illustrate their positions at the epoch of the \textit{WISE} observations, overlaid on a WIRCam \Ks-band image that shows both objects distinctively (a slight mismatch can be seen in the position of GU Psc b since the WIRCam image was taken about 2 yr after \textit{WISE} observations). The plus signs give the position of the flux barycenter in $W1$ (yellow sign, closer to the contaminating galaxy) and in $W2$ (blue sign, closer to GU Psc b) fluxes in \textit{WISE} images; each symbol's size corresponds to the position uncertainty in that bandpass.}
\label{fig:composite}
\end{center}
\end{figure}

\subsubsection{Far-red Optical Photometry}
The companion was originally identified as part of a survey of young low-mass stars with Gemini-South/GMOS imaging in $i$ and $z$ bands \Artigauprep. The original three 300 s exposures in $i$ and three 200 s exposures in $z$ were taken on 2011 September 22 (see Figure \ref{fig:composite}). The custom data reduction procedure included overscan and fringe subtraction and flat-field correction. Astrometry was anchored to the USNO-B1 catalog. The images were median combined and the magnitude zero point was determined through a cross-match with the Sloan Digital Sky Survey (SDSS). Among the 91-star sample of the survey, GU Psc's companion was the only credible candidate found  for separations ranging between 5\arcsec\ and 10\arcsec\ (depending on the primary's brightness) and the edge of the GMOS 5$\farcm$5 field of view (FoV). It was detected in the $z$-band image, but not in the $i$ band. Follow-up observations with the same instrument and observational setup were made on 2011 October 18 to obtain a deeper $i$-band image: five 300 s $i$-band images were taken, as well as an additional 200 s $z$-band image. The $z$-band photometry was consistent with that of the discovery data set, confirming that this object was not a transient or artifact. The new $i$-band imaging still did not reveal the companion but provided a 3$\sigma$ upper limit on the flux of $i$ $>$ 25.28, indicating a very red $i$ -- $z$ color ($>$ 3.53, $3 \sigma$). 

\subsubsection{Near-infrared Photometry and Astrometry}
Follow-up NIR photometry was carried at the CFHT with WIRCam \citep{Puget2004}. GU Psc b was first observed in the $J$ band on 2011 October 10 for a total integration time of 14.2 minutes with single exposures of 50\,s (see Figure \ref{fig:composite}). The target was centered on the North-East WIRCam detector and observed using a large dither pattern (15 positions or more) with the nominal 60\arcsec\ amplitude. The images were preprocessed by the IDL Interpretor of WIRCam Images ($`$I$`$iwi)\footnote{\url{www.cfht.hawaii.edu/Instruments/Imaging/WIRCam}\\ \url{/IiwiVersion1Doc.html}} which performs the dark subtraction, flat fielding, bad pixel masking, and sky subtraction. The final stacks were produced with SExtractor, SCAMP, and SWarp \citep{BertinArnouts1996,Bertin2006,Bertin2002} and the zero point was determined using color-corrected 2MASS photometry converted to the MKO system with \citet{Leggett2006} color transformations.

NIR photometry follow-up was also made on 2011 November 5 at the 1.6\,m telescope of the Observatoire du Mont-M\'{e}gantic, with the NIR camera CPAPIR \citep{Artigau2004} in queue mode \citep{Artigau2010}. A set of 270 \Ks band images, each with two 10.1\,s coadds were taken for a total exposure time of 91 minutes. A standard image processing (same pipeline as described in \citealt{Artigau2011}) was performed and yielded \Ks\ = 17.10\plmo0.15 for the object. The resulting \Ks-band versus \textit{Wide-field Infrared Survey Explorer} (\textit{WISE}) colors (see the following section, Section \ref{subsec:obsredWISE}) suggested a T dwarf spectral type and prompted both additional photometric observations at the CFHT/WIRCam and spectroscopic follow-up with Gemini-North/GNIRS.

The second and third epoch of photometry with WIRCam in $J$ were thus acquired on 2011 December 26, 28, and 29 and on 2012 September 7. Images in $Y$, $H$, and \Ks\ were also acquired on 2012 September 7. The observation strategy was the same as the one explained above, with total integration times of 45.8 and 30.4 minutes, respectively, in the two $J$-band epochs and 30.4, 19.0 and 8.3 minutes for the $Y$, $H$ and \Ks\ stacks, respectively. Single exposures were 150\,s, 50\,s, 15\,s, and 25\,s for $Y$, $J$, $H$, and \Ks, respectively. The $Y$-band zero point was determined through the observation of a spectrophotometric standard. The 2011 October and 2012 September $J$-band images allowed precise multi-epoch astrometric measurements: a linear astrometric solution was determined for each based on the 2MASS point source catalog \citep{Cutri2003}. 

\subsubsection{WISE Photometry}
\label{subsec:obsredWISE}
In the \textit{WISE} All-Sky Source Catalog\footnote{Available at \url{wise2.ipac.caltech.edu/docs/release/allsky}.} \citep{Cutri2012}, there is a source detected at the position of the companion with $W1$ = 15.818\plmo0.064 and $W2$ = 15.039\plmo0.120. However, the WIRCam, GMOS, and CPAPIR images also show a faint extended object, most likely an edge-on spiral galaxy, \env3\arcsec\, south east of GU Psc b. The inset of Figure\,\ref{fig:composite} shows the position of GU Psc b and of the galaxy at the epoch of the \textit{WISE} measurements (red circles), overlaid on the WIRCam \Ks-band image. The position of GU Psc b was deduced from WIRCam images and the proper motion in Table \ref{table:prop}. The yellow and cyan plus signs show the \textit{WISE} $W1$ and $W2$ fluxes barycenters, respectively. They both fall within $1\sigma$ on the line joining GU Psc b and the interloping galaxy, which confirms that the \textit{WISE} photometry is a blend of the two objects. The position along the line allows one to derive the relative contribution of each object to the blended flux of the catalog. The galaxy contributes 71\%\plmo10\% of the $W1$ flux and 46\%\plmo10\% of the $W2$ flux. 
The resulting photometry for GU Psc b, listed in Table \ref{table:prop}, is overall consistent with mid-T photometry for field objects (e.g., \citealt{Kirkpatrick2011,Dupuy2012}).

\subsubsection{Near-infrared Spectroscopy}
The spectrograph GNIRS on Gemini-North was used in its cross-dispersed mode to obtain a NIR 0.9--2.4\,\micron\ spectrum with a resolving power of \textit{R} \env 800. We used the 0\farcs675 slit, the short (0\farcs15 pixel$^{-1}$) camera and the 31.7\,l mm$^{-1}$ grating. The target observations, acquired on 2012 November 14, were followed by the observation of an A0 spectroscopic standard star (HIP13917) to calibrate the flux and correct for telluric absorption lines. A total of 50 minutes of observation was taken, subdivided in ten 300\,s exposures. The reduction was carried with the pipeline presented in \citet{Delorme2008b} and \citet{Albert2011}. Wavelength calibration was made using bright OH sky lines \citep{Rousselot2000}. To perform the flux calibration, the spectrum was integrated over the WIRCam $Y$, $J$, $H$, and \Ks\ bandpasses, respectively\footnote{Using the transmission curves available at \\ \url{www.cfht.hawaii.edu/Instruments/Filters/wircam.html}.}. For each filter, we then evaluated the factor by which one must multiply the integrated flux to get the WIRCam photometric measurements. A linear fit of the factors was used to rectify every wavelength of the spectrum. The calibrated spectrum is shown in Figures \ref{fig:specstand}, \ref{fig:specmodelab}, and \ref{fig:specmodelc}, along with photometric points outside the range covered by the spectrum ($z$, $W1$, and $W2$).

\begin{figure*}
\begin{center}
\includegraphics[width=7.5cm]{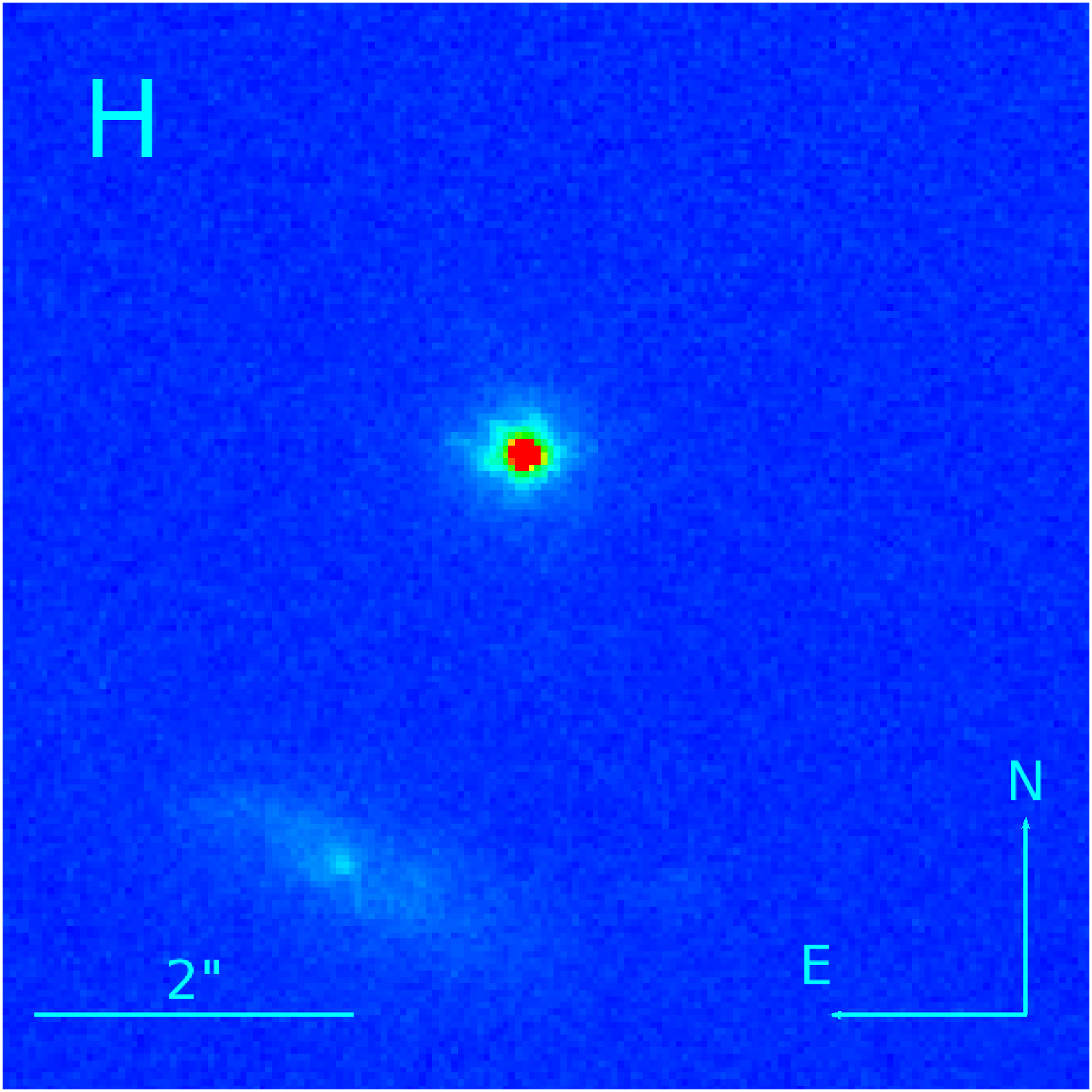}
\includegraphics[width=7.5cm]{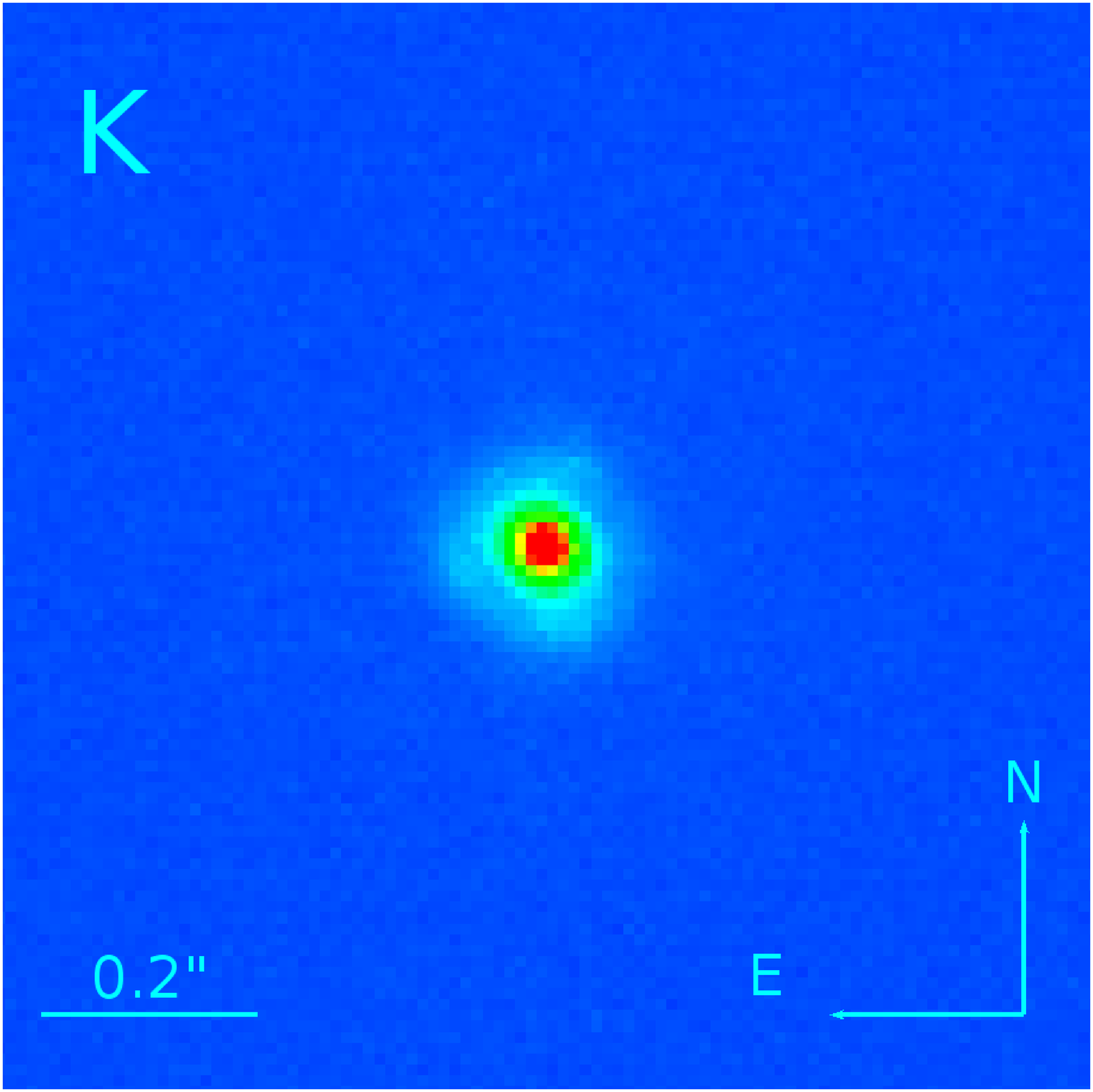}
\caption{LGS AO $H$- and $K$-band images of the GU Psc b. Left: the $H$-band image shown is 6\farcs8 wide, the galaxy present in the inset of Figure \ref{fig:composite} can be seen in the lower left corner. Right: the $K$-band image shown here is \env1\farcs7 on a side. There is no other point source detected in the FOV (10\farcs2 $\times$ 10\farcs2).}
\label{fig:Keck}
\end{center}
\end{figure*}

\subsubsection{High-resolution Near-infrared Imaging}
On 2013 November 19, we used the sodium Laser Guide Star (LGS) AO system of the Keck II Telescope \citep{Wizinowich2006, vanDam2006}, located on the summit of Mauna Kea, in Hawaii, to verify if the companion is a resolved binary. The LGS constitutes the reference wavefront for the AO correction, and the tip-tilt was monitored using the close (26\farcs4 away) star SDSS J011237.05+170456.9 ($r_{\rm{AB}}$ = 15.742\plmo0.004; SDSS DR 9; \citealt{Ahn2012}). We used the near-infrared camera NIRC2 in its narrow mode (FoV 10\arcsec $\times$ 10\arcsec, pixel scale of 9.942\,mas pixel$^{-1}$) for the $K$-band images and in its wide mode (FoV 40\arcsec $\times$ 40\arcsec, pixel scale of 39.686\,mas pixel$^{-1}$) for the $H$-band images. We obtained 12 images of two coadditions of 60\,s in $K$ band, for a total integration time of 24 minutes, and nine images of four coadditions of 30\,s in $H$ band, for a total integration time of 18 minutes. The images were obtained using a three-point dither pattern that avoided the noisy quadrant in the lower-left portion of the array. The positional offset between images varied between 1\farcs5 and 3\arcsec. Conditions were photometric during the observation. 
Standard reduction techniques were used: images were divided by the dome flat and a median sky image constructed from all the observations of the night was subtracted. Then every image was shifted and stacked to produce a final image in each band (see Figure \ref{fig:Keck}).

\section{Results}
\label{sec:res}

\subsection{Physical Properties of the Host Star}
\subsubsection{GU Psc: A Young Low-mass Star Candidate Member of the AB Doradus Moving Group}
\label{sec:GUPsc}
GU Psc was originally identified as a highly probable M dwarf by \citet{Zickgraf2003}. Then, \citet{Riaz20062} identified it as an M3\plmo0.5 using the TiO5 index \citep{Reid1995} and obtained a visible spectrum that allowed the measurement of its \Ha\ emission line equivalent width (EW), which is a proxy for its chromospheric activity, and thus youth. 

Through Bayesian inference, GU Psc was then identified by \citet{Malo2013} as a highly probable (98\%) member of the ABDMG. The Bayesian analysis makes use of a priori knowledge of known associations (Galactic position, space velocity, and $I_{C}-J$ absolute photometry) and compares these properties with observables of a given candidate (sky position, proper motion, and $I_{C}$ and $J$ apparent magnitudes). The analysis gives as an output the membership probability of the star for every association considered as well as the probability that it is a field star unrelated to these associations. It also gives the most probable radial velocity the candidate should have if it were a true member of a given association (accurate to a few \kms) and the most probable statistical distance \ds\ it would have. \citet{Malo2013} showed that this statistical distance agrees with the trigonometric distance within \env10\% for bona fide members of the associations. For GU Psc, using the average proper motion shown in Table\,\ref{table:prop}, they derived \ds\ = 48\plmo5\,pc, which is the value we adopt for the distance hereafter and predicted a \vrad\ = --1.5\plmo1.9\,\kms. Using any of the proper motion measurements available in the literature does not change these results significantly. As part of a comprehensive radial velocity follow-up program (\Maloprepalt), multi-epoch radial velocity measurements of GU Psc through NIR and optical spectroscopy were secured. 

The results, summarized in Table \ref{table:vrad}, yield a weighted average radial velocity of $<$\vrad$>$ = --1.6\plmo 0.4 \,\kms, in excellent agreement with the predicted radial velocity for membership in ABDMG. Adding the radial velocity observable to the Bayesian analysis yields an increased membership probability of 99.9\% for ABDMG. These observations also show that GU Psc is a relatively fast rotator with a \vsini\ of 23\,\kms. 

Note that, as mentioned in \Gagne, the probabilities mentioned here should not be interpreted as absolute. Even though any given star is a priori less likely to belong to a given association than to the field (there are much less stars in the association than in the field), the precise values of these prior probabilities are very uncertain so all hypotheses are considered as equally likely in the analysis of \citet{Malo2013}. They estimate that for the candidates of ABDMG with a membership probability over 90\%, the contamination rate (false positive) is about 14\%.   

We also used the analysis of \Gagne\ to evaluate GU Psc's membership. This analysis differs from that of \citet{Malo2013} in two major aspects. First, it uses a different method to outline the three-dimensional regions covered by the bona fide members of an association in the Galactic position space $X$, $Y$, $Z$ and in the Galactic velocity space $U$, $V$, $W$. While \citet{Malo2013} uses ellipsoids with the three axes aligned on the Galactic coordinate system, \Gagne\ use a more realistic approach where the ellipsoids can be aligned in any direction. Second, \Gagne\ uses the knowledge of the distance of known members of a given association as an additional prior on the plausible distance that a candidate can have. In agreement with \citet{Malo2013}, this analysis points toward a membership in ABDMG for GU Psc, albeit with a smaller probability (88$\%$). It finds a compatible statistical distance of 47\plmo5\,pc and a very similar predicted radial velocity (--1.8\plmo2.0\,\kms). It also yields a non-negligible probability of 12\% for the membership in the younger (12--22\,Myr) $\beta$ Pictoris Moving Group (\BPMG), associated with a smaller statistical distance, \ds\ = 32\plmo3\,pc. 

Thus, both analyses suggest that GU Psc is a member of a young association, either ABDMG or \BPMG, with a much higher probability for the former. 

\begin{table}[htbp]
 \newcolumntype{A}[1]{>{\arraybackslash}m{#1}} 
 \newcolumntype{W}[1]{>{\centering\arraybackslash}m{#1}} 
\begin{center}
\caption{Radial Velocity and Projected Rotational Velocity of GU Psc A}
\label{table:vrad}
\begin{tabular}{W{1.2cm}W{1.6cm}W{1.8cm}W{1.8cm}W{0.5cm}} 
\hline \hline      
                   & Date &\vrad\                   & \vsini\                   & Note\\
                   &       &(\kms)       & (\kms)        & \\
\hline
&&&&\\
Measured & 2010 Nov 19         & 0.4\plmo1.2 & 25.4\plmo2.6 &a\\ 
                   & 2012 Jul 11            & --1.8\plmo0.7 & 23.1\plmo2.3 &b\\
                   & 2013 Jul 21            & --1.7\plmo0.7 & 24.1\plmo2.3 &b\\
                   & 2012 Jan 6           & --1.6\plmo0.3 & 22.5\plmo0.9 &c\\
                   &  Average            & --1.6\plmo 0.4 & 23.0\plmo 1.4&d \\
&&&&\\
Predicted  &                    & --1.5\plmo1.9\tnme    &  &\\
\hline
&&&&\\
\multicolumn{5}{A{8.3cm}} {\textbf{Notes}}\\ 
\end{tabular}
\begin{tabular}{p{0.1cm}p{8.1cm}}
\tnma & Gemini-S/PHOENIX NIR spectroscopy.\\
\tnmb & VLT/CRIRES NIR spectroscopy.\\
\tnmc & ESPaDOnS/CFHT optical spectroscopy.\\
\tnmd & Weighted average of the 4 measures.\\
\tnme & Bayesian analysis \citep{Malo2013}.\\
\end{tabular}
\end{center}
\end{table}

\subsubsection{The Age of the AB Doradus Moving Group}
The age of ABDMG, first identified as a moving group by \citet{Zuckerman2004}, is subject to debates. The comparative analysis of ABDMG and the open cluster IC 2391 in an $M_V$ versus $V$ -- $K$ color--magnitude diagram led \citet{Luhman2005} to derive an age between 75 and 150\,Myr, roughly coeval with the Pleiades (for which they adopted an age of 100--125\,Myr). Using an $M_{I}$ versus $V$-$I$ diagram and the lithium EW, \citet{LopezSantiago2006} formulated the hypothesis that the group could be composed of two subgroups: one younger (30--50\,Myr) and one older (80--120\,Myr). The Li EW was also used in two other studies to deduce a lower limit on the age of 45\,Myr \citep{Mentuch2008} and an age of 70\,Myr  \citep{daSilva2009}. Recently, \citet{Barenfeld2013} studied the kinematics and the abundance of 10 different elements in 10 members of the ``stream" (i.e., stars that are not among the nine stars considered as the ``nucleus" by \citealt{Zuckerman2004}) and pointed out that many stars traditionally associated with ABDMG do not have a similar chemical composition and/or were not likely formed at the same position as the ABDMG nucleus. They concluded that the remaining members are at least 110\,Myr, based on the fact that the group still has zero-age main sequence K stars members. Considering all these studies, we adopt a conservative age of 100\plmo30\,Myr for ABDMG as a whole.

\subsubsection{Youth Indicators}
To better constrain the age of GU Psc, we consider here other age indicators. Table\,\ref{table:age} summarizes all the information on the age of the GU Psc system. 

Over time, the coronal activity that is induced by magnetic field is reduced, which causes the X-ray emission to decrease \citep{PreibischFeigelson2005}. Using GU Psc's X-ray count rate and hardness ratio (HR1) measured by \textit{ROSAT} (\citealt{Voges1999}; see Table \ref{table:prop}) in the relation given in \citet{Schmitt1995} yields a value of 4.90 $\times$ $10^{-13}$\,erg s$^{-1}$ cm$^{-2}$ for the X-ray flux, thus an X-ray luminosity of \logLx\ = 29.1\plmo0.3\,erg \smun\ at \ds\ = 48\,\plmo\,5\,pc. This X-ray luminosity is very similar to that of single low-mass ABDMG members; if we use a similar procedure to evaluate the \logLx\ of the six bona fide M dwarfs members listed in \citet{Malo2013}, we obtain \logLx\ = 29.03\,erg \smun, with a dispersion of 0.07 dex. The X-ray luminosity of GU Psc is also consistent with that of other candidate members of ABDMG, such as the M3.5 star J01225093$-$2439505 (hereafter 2M0122) that has \logLx\ = 28.7\plmo0.2\,erg \smun\ \citep{Bowler2013}, or the M2 star J235133.3$+$312720, that has a \logLx\ = 29.3\plmo0.2\,erg \smun\ \citep{Bowler2012}. GU Psc's X-ray luminosity is however significantly higher than that of field stars of similar mass. For example, the \logLx\ of the 42 single field M dwarfs listed in \Maloprep\ has a mean of \logLx\ = 27.6\,erg \smun\ with a dispersion of 0.5 dex. 
At the statistical distance for the \BPMG\ given by \Gagne\ analysis (\ds\ = 32\plmo3\,pc), GU Psc's X-ray luminosity would be \logLx\ = 28.8\plmo0.3\,erg \smun, which is also not consistent with that of \BPMG\ members: the mean value computed for the nine single bona fide M dwarfs of the \BPMG\ in \citet{Malo2013} is \logLx\ = 29.63\,erg \smun, with a dispersion of 0.16 dex. 
The X-ray activity thus favors an ABPMG membership and not a \BPMG\ membership.

\begin{table}[h]
\newcolumntype{W}[1]{>{\arraybackslash}m{#1}} 
\begin{center}
\caption{Age Estimate of the GU Psc System}
\label{table:age}
\begin{tabular}{W{4.0cm}W{4.0cm}} 
\hline \hline      
Method    &Range of Age Allowed\\
\hline
&\\
Bayesian analysis (kinematic \& photometry) & ABDMG candidate member (100\plmo30\,Myr)\\
 \hline 
&\\
X-ray emission & Similar to ABDMG members\\ 
\Ha\ emission EW & $<$ 2\plmo0.5\,Gyr\\
\Ha\ emission W10\% & $>$ 10\,Myr\\
No infrared excess & $>$ 10\,Myr\\
Li absorption absent & $>$ 22\,Myr\\
Rotation period & If $M_{\star}$ $>$ 0.35\,\Msun, $<$ 650\,Myr\\
&If $M_{\star}$ $<$ 0.35\,\Msun, no constraint\\  
\hline
&\\
Adopted & 100\plmo30\,Myr\\
\hline
\end{tabular}
\end{center}
\end{table}

\begin{figure}[htbp]
\begin{center}
\includegraphics[width=8.5cm]{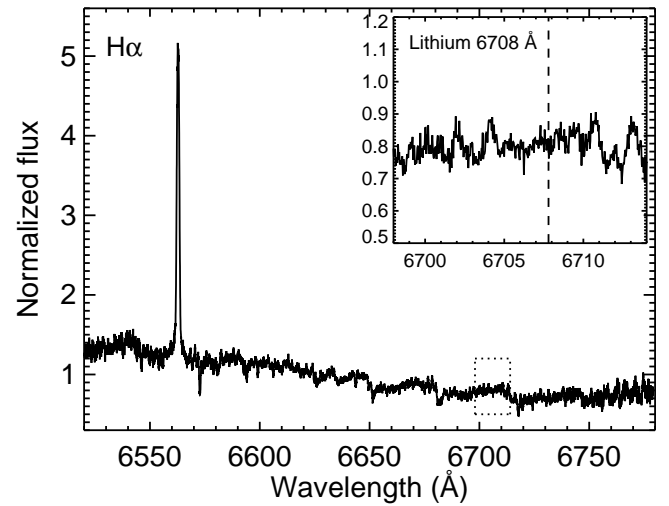}
\caption{Host star ESPaDOnS optical spectrum between 6520\,\AA\ and 6780\,\AA. The H$\alpha$ emission line at 6562.8\,\AA\ is clearly seen. The equivalent width of the band is --3.96\,\AA. The 6708\,\AA\ lithium absorption line is however not detected, we measure a 3$\sigma$ upper limit of EW$_{\rm{Li}}$\lta18 m\AA.} 
\label{fig:specopt}
\end{center}
\end{figure}

The reduction of the magnetic activity occurring as the star evolves is also traceable by the diminution of the \Ha\ emission line at 6562.8\,\AA. In our visible spectrum (Figure \ref{fig:specopt}(a)), we measure \EWHa\ = $-3.96$ \AA. According to \citet{West2008}, the activity lifetime of an M3 is 2\plmo0.5\,Gyr, thus the presence of \Ha\ in emission implies that GU Psc is likely younger than this. Also, according to the criteria developed in \citet{White2003}, the 10\% width of the same \Ha\ emission line (W10\% = 125\,\kms) is consistent with a star that is non accreting (W10\% $<$ 270\,\kms), thus older than \env10\,Myr \citep{BarradoyNavascues2003}. This lower limit is also consistent with the fact that no disk is seen in the form of a mid-infrared excess. Indeed, when the $J$, $H$, \Ks\ bands are fitted with a model spectrum (BT-Settl AGSS2009; \citealt{Allard2012}), the \textit{WISE} photometry in the four bands falls directly on the model spectral energy distribution (SED). These three indicators are consistent with a membership in either ABDMG or \BPMG.

The absence of the lithium absorption line at 6708\,\AA in the optical spectrum of GU Psc (EW$_{\rm{Li}}$\lta18 m\AA, 3$\sigma$; see Figure \ref{fig:specopt}(b)) also yields a minimum age for GU Psc. Indeed, the early M members of ABDMG show lithium measurements compatible with this upper limit \citep{Mentuch2008,Yee2010}. Our upper limit is also compatible with the wide range of lithium absorption that the low-mass stars show at the younger age (12--22\,Myr) of \BPMG, which varies from an upper limit of a few tenths of m\AA\ up to \env500\,m\AA\  \citep{Mentuch2008,Blinks2013arxiv}. Indeed, Figure 5 of \citet{Mentuch2008} shows that the youngest association for which early M stars show little or no lithium is \BPMG; there is no early M star with EW$_{\rm{Li}}$\lta350\,\AA\ in the TW Hydrae Association (TWA) nor in the $\eta$ Chamaelontis Cluster, which are both \lta20\,Myr (e.g., \citealt{Fernandez2008}). We can thus estimate that GU Psc is likely older than \env22\,Myr.

For an object with a mass above the fully convective limit (\gta0.35\,\Msun, \citealt{Chabrier1997}), the rotation rate increases as the star contracts toward the main sequence, reaches a plateau at an age comparable to that of ABDMG, and then slows down due to various interactions \citep{Sills2000}. The rotation period can thus help to constrain the age. According to the SuperWASP photometric survey, GU Psc is a variable star with a 1.0362\plmo0.0005 day period \citep{Norton2007}. That could be indicative of a relatively fast rotator, which is also suggested by its relatively large \vsini\ of 23\,\kms. If GU Psc is not fully convective ($>$ 0.35\,\Msun), the \env1 day rotation period suggests an upper limit on the age of \env650\,Myr (see Figure 12 in \citealt{Irwin20112}). With the $I_{C}-J$ listed in Table \ref{table:prop} and the age of ABDMG (100\plmo30\,Myr), GU Psc's mass $M_{\star}$ is estimated to be between 0.30 and 0.35\,\Msun\ (using the models of \citealt{Baraffe1998}), which is very close to the limit for a star to be fully convective. Without a parallax, it is challenging to determine whether or not GU Psc has a fully convective structure. If GU Psc is fully convective, the upper limit of the age we can set using the rotation period is much greater, since the spin-down time of fully convective objects is very long ($>$ 5\,Gyr; \citealt{Irwin20112}).

It is interesting that HIP17695, the only single bona fide member of ABDMG with a spectral type similar to GU Psc (M2.5; \citealt{Malo2013}), has a \vsini\ (18\,\kms; \citealt{daSilva2009}) and rotation period (\Prot\ = 3.87 days; \citealt{Messina2010}) that are close to those of GU Psc. This object is probably at the low-age end of ABDMG (\env70\,Myr), given its X-ray luminosity, \Ha\ and Li EW \citep{Zuckerman2004}. 

\vspace{3mm}
Considering that the Bayesian analysis favors a membership in ABDMG, and that other youth indicators suggest an age consistent with that association, we adopt hereinafter the age of ABDMG (100\plmo30\,Myr) and the associated statistical distance (\ds\ = 48\plmo5\,pc) for GU Psc's system.

\subsubsection{Metallicity}
We evaluated the metallicity of GU Psc using two metallicity calibrations developed recently, specifically for M dwarfs.  One must bear in mind however that these calibrations were developed for field stars and are not necessarily appropriate for young stars. \citet{Newton2014} calibration is based on the Na line at 2.2\,\micron. For GU Psc, we obtain [Fe/H]= +0.10\plmo0.13. \citet{Mann2013} improves previous calibrations (notably \citealt{Terrien2012} and \citealt{RojasAyala2012}) and presents metallicity calibrations based on various features in optical and NIR. We obtained [Fe/H]$_{H}$ = --0.14\plmo0.09 with the $H$-band calibration and [Fe/H]$_{K}$ = 0.04\plmo0.08 with the $K$-band calibration\footnote{See the IDL program available online at \\ \url{https://github.com/awmann/metal}.}. 
While the value derived using the $H$-band is slightly lower than the others, the ones derived with \citet{Mann2013} and \citet{Newton2014} using $K$ band are consistent, within uncertainties, with each other and with the one derived by \citet{Barenfeld2013} for 10 ABDMG members, [Fe/H] = +0.02\plmo0.02. 

\subsubsection{Constraints on Multiplicity of the Host Star}
\begin{figure}[htbp]
\begin{center}
\includegraphics[width=8.5cm]{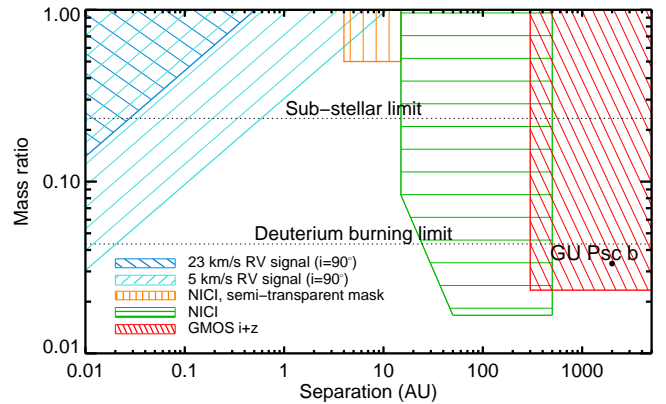}
\caption{Constraints on the presence of another companion around GU Psc A. The shape of the rotation-broadened line profile in the CRIRES and ESPADONS spectra exclude the blue region (\vsini\ $>$ 23\,\kms). The multi-epoch radial velocity measurements are stable to within 1\,\kms\ (see Table \ref{table:vrad}) and match that expected for the ABDMG along the line-of-sight to within 1\,\kms. This conservatively rules out SB1 binaries with a radial velocity semi-amplitude \vsini\ $>$ 5\,\kms\ (cyan area). The spectroscopic constraints are shown for systems with inclinations close to edge-on ($i\sim90$\deg). At separations between a few AUs and a few hundred of AUs, NICI high-contrast imaging allows ruling out companions, down to the planetary-mass regime for the largest separations (green and orange area). At separations greater than a few hundred AUs, GMOS imaging eliminates the possibility of another companion down to planetary-mass regime (red region). } 
\label{fig:multiplicity}
\end{center}
\end{figure}

The high-contrast imaging and high-resolution spectroscopy observations we made on the primary provide strong constraints on the mass ratio and separation of a possible planetary-mass, brown dwarf, or stellar companion (see Figure \ref{fig:multiplicity}). 

First, high-resolution spectroscopy shows a single-line profile. Considering the measured projected rotational velocity of GU Psc (\vsini\ \env23\,\kms; Table \ref{table:vrad}), this excludes a double-line binary (SB2) with components showing $\Delta$\vsini\,$>$ 23\,\kms. 
Second, the multi-epoch radial velocity measurements obtained with ESPaDOnS and CRIRES (Table \ref{table:vrad}) are stable at the \kms\ level and consistent with the value predicted for an ABDMG member in GU Psc's line of sight within $<$ 1\,\kms, which excludes many cases of single-line binary (SB1). 
We adopt a conservative upper limit on GU Psc's radial velocity semi-amplitude of \vsini\ $<$ 5\,\kms, since a larger discrepancy between the measured radial velocity and the one predicted for GU Psc in ABDMG would correspond to a $3\sigma$ outlier.
Note however that for both spectroscopic constraints (\vsini $>$ 23\,\kms\ and \vsini\ $>$ 5\,\kms), a near-equal luminosity binary system would not be detected for nearly pole-on geometries. In the case of the second constraint (\vsini\ $>$ 5\,\kms), a companion with the same luminosity as GU Psc would not be detected either. The blue and cyan regions of Figure \ref{fig:multiplicity} show the mass ratio and separation ranges excluded by these two constraints, respectively, for an orbit orientation of $i$\env90\deg.

At greater separations, no companions (other than GU Psc b) were found through imaging observations with NICI or GMOS. Inside NICI's semi-transparent mask, which absorbs \env6\,mag in the central 0\farcs32 radius region (15\,AU at \ds\ = 48\,pc), no object is seen at separations greater than one FWHM of the point-spread function (0\farcs08, or 4\,AU), down to a flux ratio of 4 (1.5\,mag). 
The NICI data were taken with the 4\% \CHq\ on and off filters within the $H$ band; a companion 1.5\,mag fainter than GU Psc would have $M_H$ = 9.3 which corresponds to a \env M6 spectral type \citep{Dupuy2012} and a temperature 400\,K cooler than GU Psc  (i.e., respectively \env3300\,K and \env2900\,K, the difference being more accurate than the absolute values, see Figure 5 in \citealt{Rajpurohit2013}). Using the BT-Settl model (AGSS2009, \citealt{Allard2012}) at an age of 100\,Myr, a difference of $400$\,K for a primary star in the 3000--3500\,K range leads to a maximal mass ratio of 32\%--54\%. The orange region of Figure \ref{fig:multiplicity} shows the excluded region from this observation, adopting 50\% as a conservative upper limit on the mass ratio. 
NICI imaging beyond the edge of the mask shows no background companion out to a separation of 9\arcsec (430\,AU) at a $5\sigma$ contrast of $\Delta H\sim12$, yielding an upper limit of $M_H\sim18$, or a mass limit of \env5\,\MJ\ (green region of Figure \ref{fig:multiplicity}). 
The GMOS imaging has a $5\sigma$ $z$-band limit at 23.2, which translates to a mass limit of \env7\,\MJ,and, thus, a mass ratio of 2.3\% (red region of Figure \ref{fig:multiplicity}). 

\citet{Malo2013} mentioned the possibility that GU Psc might be a binary star. Indeed, the absolute magnitude GU Psc assuming \ds\ = 48\,pc, $M_{J}$ = 6.80, is about 0.78 mag brighter than the one predicted for a single ABDMG member with GU Psc's color ($I_{C}-J$ = 1.44; see the $M_{J}$ versus $I_{C}-J$ color-magnitude diagram on Figure 3 in \citealt{Malo2013}). The magnitude dispersion along the ABDMG empirical sequence for that $I_{C}-J$ (\env0.5 mag; \citealt{Malo2013}) is important. A limited number of bona fide low-mass members are known in this association and there is probably an intrinsic age spread among the members. Thus, this over luminosity may not be significant. If it is, it could be due to an unseen companion, but it could also be attributed to other factors (for example to an important chromospheric activity; \citealt{Riedel2011}). 

In conclusion, the various data sets we obtained pose stringent constraints on the presence of another companion around GU Psc. The near-equal luminosity binary scenario can be virtually excluded at all separations, unless a rather unlikely geometry is invoked, such as a near pole-on geometries for SB1 or SB2 cases or an alignment of a companion behind the star at the time of the high-resolution imaging observations. Besides, there is still a possibility of a stellar companion with a maximum mass of about half that of GU Psc's between 1 and 10\,AU or of a brown dwarf companion inward of \env10\,AU. 

\subsection{Physical Properties of the Companion}
\subsubsection{Proper Motion}
\begin{figure}[htbp]
\includegraphics[width=8.5cm]{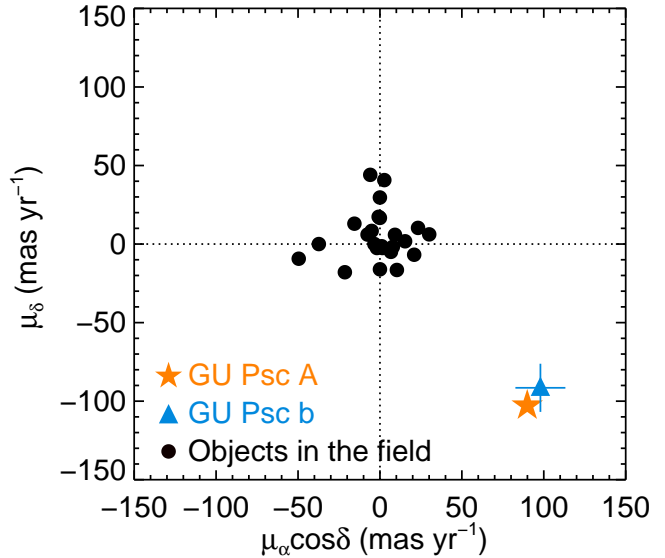}
\caption{Proper motion in declination and right ascension for GU Psc A, GU Psc b, and stars in the field computed with the 2011 October and 2012 September $J$-band WIRCam images. The primary is saturated on these images so a weighted average of the proper motions found in the literature is shown. The uncertainty on GU Psc A proper motion is smaller than the plot symbol.}
\label{fig:mvtpropre}
\end{figure}

Figure \ref{fig:mvtpropre} shows the proper motion of GU Psc and of GU Psc b. For the primary, there are several proper motion measurements reported in the literature \citep{Roeser2010,Ducourant2006,Roeser2008,Ahn2012,Schlieder2012}. Since these measurements are not independent, we adopt the mean value and use the average uncertainties as the error of the resulting mean: $\mu_{\alpha}\cos\delta$ = 90\plmo6\,mas yr$^{-1}$, $\mu_{\delta}$ = --102\plmo6\,mas yr$^{-1}$.
For the companion, we used the two $J$-band epochs taken at CFHT/WIRCam in 2011 October and 2012 September (11 months apart) and found $\mu_{\alpha}\cos\delta$ = 98\plmo15\,mas yr$^{-1}$, $\mu_{\delta}$ = --92\plmo15\,mas yr$^{-1}$. The figure also shows the proper motions of stars in the WIRCam image field, computed like that of GU Psc b. The companion proper motion is clearly consistent with that of the star and inconsistent with that of the other stars in the field, thus confirming it is gravitationally bound. 

This conclusion is strengthened by the fact that the probability of finding an un-associated T dwarf with this proper motion in the vicinity of a young M dwarf is very low.
To assess this probability, we first consider only proper motion. Using the 64 T dwarfs in the \citet{Dupuy2012} sample (Table 9) that have a parallax and proper motion with errors smaller than 10\%, we compute a median sky plane velocity, 35\,\kms, which we assume to be typical of T dwarfs. Assuming an isotropic random Gaussian distribution of velocities, this velocity corresponds to a Gaussian dispersion per coordinate of 30\,\kms, which translates back into a dispersion of 130 mas yr$^{-1}$ at 48\,pc. A numerical integration of a two-dimensional Gaussian shows that only $\sim1.5\%$ of random T dwarfs would match GU Psc's proper motion at the $<2\sigma$ level.
Then, we consider the sky position. The local density of T0--T5.5 dwarfs is $1.4^{+0.3}_{-0.2}$ $\times$ $10^{-3}$ pc$^{-3}$ \citep{Reyle2010}, there are therefore \env730 early T dwarfs within 50\,pc of the Sun. Our survey sampled a near-circular field with a 5$\farcm$5 diameter around 91 stars, covering a total area of 0.6\,deg$^{2}$ or 1.5 $\times$ $10^{-5}$ of the entire sky. There is thus only a \env1\% chance of finding a random single field early-T dwarf with properties comparable to GU Psc b within a GMOS FoV of one survey stars. 
The combined likelihood of a false positive match in both proper motion and position is therefore of the order of 2 $\times$ $10^{-4}$, and it is likely to be much lower as GU Psc and GU Psc b both display signs of youth. 

\subsubsection{Constraints on Multiplicity of the Companion}
\label{subsubsec:MultiGUb}
The $H$- and $K$-band observations of the companion made with LGS AO at Keck show only one object. The companion is thus not a resolved binary. The $K$-band image excludes, at the $>$5$\sigma$ level, any second object that would have a $\Delta$$K$ \lta 4 for separations between 0\farcs08 and 10\arcsec\ (\env 4--480\,AU at 48\,pc). Using $M_K$ \env\ 14 for GU Psc b, this $\Delta$$K$ corresponds approximately to a T8 or later, according to the polynomial relations given in \citet{Dupuy2012}. For smaller separations, the $K$-band observations exclude a companion brighter than a typical T7.5 down to a separation of 0\farcs04 (\env 2\,AU). The wider $H$-band image does not show any object (besides a few galaxies) down to a $\Delta$$H$ of \env 3.6, in a radius between 0\farcs7 and the width of the field, 40\arcsec\ (\env30--1900\,AU at 48\,AU). Considering $M_H$ = 14.3 for GU Psc b, this excludes a companion earlier than \env T7.5 in that region. 

GU Psc b could still be a very tight binary object, but these observations largely exclude a T5--T6 companion in a wide range of distances and down to the typical separations of T dwarfs binaries (e.g., SDSS J153417.05+161546.1, a T1.5+T5.5 with a separation of 3.96\,AU; \citealt{Liu2006}, SDSS J102109.69--030420.1, J1021 hereinafter, a T1+T5 with a separation of 5\,AU; \citealt{Burgasser2006b}, or $\epsilon$ Indi B ab, a T1+T6 separated by 2.65\,AU; \citealt{McCaughrean2004}).

\subsubsection{Spectral Type and Spectral Characteristics} 
\label{subsubsec:Sptype}
\begin{figure*}[htbp]
\begin{center}
\includegraphics[width=12cm]{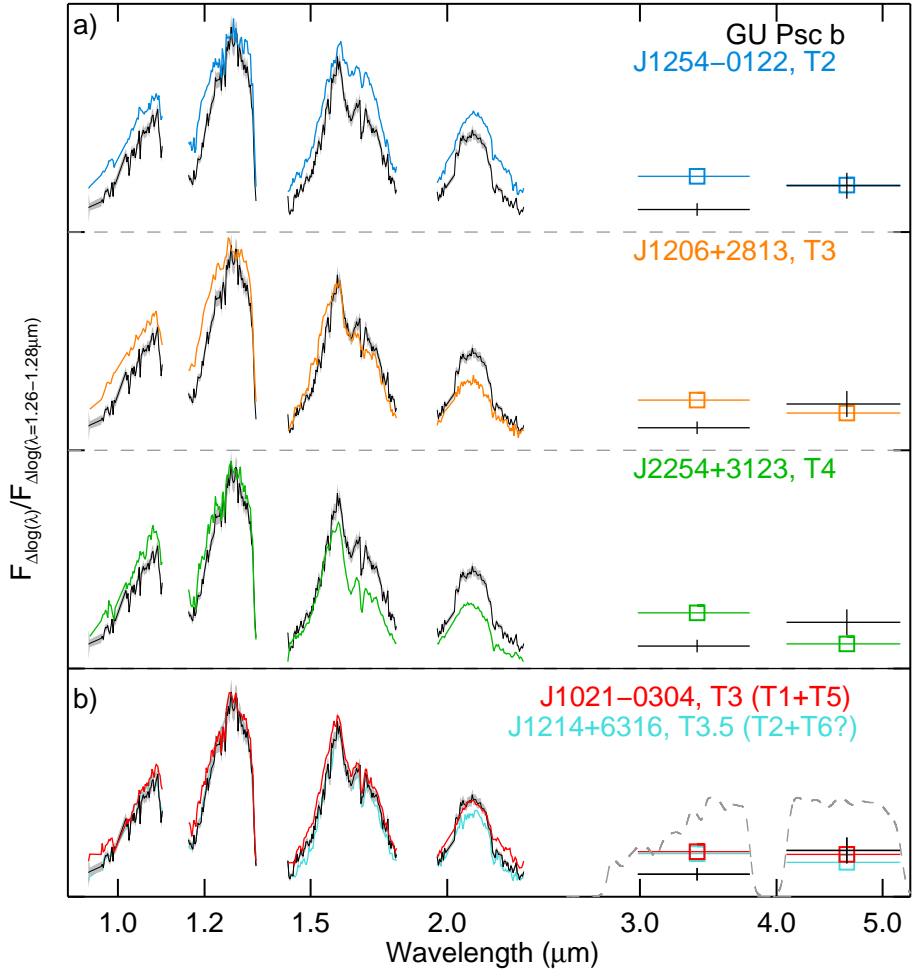}
\caption{ (a)  The GNIRS NIR spectrum and \textit{WISE} magnitudes for GU Psc b and the T2, T3, and T4 spectral standards. All SEDs are normalized at the $J$-band peak, between 1.26 and 1.28\,\micron. GU Psc b spectrum has been median-smoothed to a resolution $\lambda/\Delta\lambda\sim400$. (b) GU Psc b SED, now compared to J121440.95+631643.4, a T3.5 that is likely a binary T2+T6, according to the template fitting analysis of \citet{Geissler2011}, and J102109.69--030420.1AB (J1021), the former T3 standard that was confirmed to be a binary T1+T5 by \citet{Burgasser2006b}. GU Psc SED is much closer to that of these objects than to the SED of the standards.}
\label{fig:specstand}
\end{center}
\end{figure*}

Figure \ref{fig:specstand}(a) shows the NIR spectrum and $W1$ and $W2$ photometry of GU Psc b compared with other objects. The upper panel of the figure shows the spectral standards T2, T3, and T4. The GU Psc b spectrum was smoothed over eight points to a final resolution of $\lambda/\Delta\lambda\sim400$ to ease comparison. The spectra are from the L and T dwarf data archive\footnote{\url{staff.gemini.edu/~sleggett/LTdata.html}}. The T2 and T4 (J125453.90--012247.4, hereafter J1254, and J225418.92+312349.8) are the standards identified in \citet{Burgasser2006}. J120602.51+281328.7 was used as the T3 standard instead of J120956.13--100400.8, which was found to be a binary by \citet{Liu2010}. All spectra are normalized to their value at the peak of the $J$ band and are offset vertically for clarity. 

The global comparison of the SEDs suggests that GU Psc b is of a spectral type between T2 and T4. The absorption of \CHq\ in $H$ band is intermediate between that of a T2 and a T4, close to a T3, while the blue side of the $H$ band is better reproduced by the T4 standard. The $Y$- and $J$-band flux are also closer to the T4, even though in both cases the side of each peak is slightly underluminous. The $K$ band is not similar to any of the standards and is clearly brighter than the average T3 flux. That is probably explained by the collision-induced absorption (CIA) by \Hd\ that affects this region of the spectrum; CIA is expected to be reduced for objects with a lower gravity and/or greater metallicity \citep{Saumon2012}. The $W1$ flux is lower for GU Psc b than for any of the standards, and the $W2$ flux is stronger than the T4, closer to the T2. 

\begin{table}
 \newcolumntype{A}[1]{>{\arraybackslash}m{#1}} 
 \newcolumntype{W}[1]{>{\centering\arraybackslash}m{#1}}  
 \caption{Spectral Indices for GU Psc b}
\begin{center}  
\begin{tabular}{W{0.9cm}@{\hspace{6mm}}W{1.8cm}@{\hspace{2mm}}W{1.7cm}@{\hspace{3mm}}W{1.6cm}@{\hspace{2mm}}W{0.8cm}}
\hline\hline
Index & Numerator Range\tnma & Denominator Range\tnma & Value &Spectral Type\tnmb\\
 &({\micron})&({\micron})&&\\
 \hline\\
 \end{tabular}
\begin{tabular}{ccccc} 
\HdO$-J$  & 1.140--1.165&1.260--1.285 &  $0.30\pm0.01$   & T5 \\[3mm]
\CHq$-J$  & 1.315--1.340&1.260--1.285   & $0.491\pm0.009$   & T4.5 \\[3mm]
$W_{J}\tnmc $ & 1.180--1.230&1.260--1.285   & $0.572\pm0.009$  & \\[3mm] 
\HdO$-H$  & 1.480--1.520&1.560--1.600 & $0.400\pm0.008$   & T4 \\[3mm]
\CHq$-H$  & 1.635--1.675&1.560--1.600  &  $0.73\pm0.01$   & T3\\[3mm]
\HdO$-K$  & 1.975--1.995&2.08--2.100    & $0.46\pm0.03$   & \\[3mm]
\CHq$-K$  & 2.215--2.255&2.08--2.120   &  $0.404\pm0.009$   & T3.5 \\[3mm]
$K/J$   & 2.060--2.100&1.250--1.290   &  $0.317\pm0.006$   & \\[3mm] 
\hline
\multicolumn{5}{A{8.3cm}} {}\\ 
\multicolumn{5}{A{8.3cm}} {\textbf{Notes}}\\ 
\end{tabular}
\begin{tabular}{p{0.05cm}p{8.1cm}}
\tnma & The indices are defined as $\int_{a}^{b}{f(\lambda)d\lambda}/\int_{c}^{d}{f(\lambda)d\lambda}$, the numerator being the integrated flux in a region inside an absorption feature, the denominator being the integrated flux in an adjacent pseudo continuum.\\
\tnmb & The associated spectral type according to \citet{Burgasser2006}.\\
\tnmc & For this index, the denominator is multiplied by 2 to compensate for the larger range of the numerator.\\
\end{tabular}
\end{center}
\label{table:specindex}
\end{table}

We used the GU Psc b spectrum to compute the spectral indices defined in \citet{Burgasser2006} and establish the spectral type. The position of the wavelength ranges used and the values derived for the indices are listed in Table \ref{table:specindex}, along with the associated spectral types for the five indices for which a quantitative scale exists.
The $K/J$ index defined by \citet{Burgasser2006} measures the flux ratio between the $K$ and the $J$ band to evaluate the strength of the CIA-\Hd\ feature that is known to be sensitive to surface gravity. It has no quantitative scale, but the $K/J$ of GU Psc b is stronger than for a typical T3.5, closer to that of a T2.5 (see $K/J$ versus \CHq$-H$ diagram shown on Figure 10 of \citealt{Burgasser2006}). The $W_{J}$ index quantifies the width of the $J$-band peak. The value of GU Psc b is similar to that we computed for the T4--T5 objects present in the L and T dwarf data archive. The resulting spectral types vary between T3 and T5, with an average spectral type of \env T4. 

Altogether, the comparison of the spectrum and the indices suggest that GU Psc b is a T3.5\plmo1 spectral type, with indication of low gravity, and/or of high metallicity.

Even though GU Psc b is not a resolved binary object according to Keck NIRC2 observations (see Section \ref{subsubsec:MultiGUb}), the values of its indices, within uncertainties, satisfy two to three of the six spectral index selection criteria developed by \citet{Burgasser2010b} to identify binary systems. GU Psc b would thus qualify as a candidate binary in that scheme (weak candidates satisfy two criteria, and strong ones satisfy three criteria or more). Indeed, as shown on Figure \ref{fig:specstand}(b), the GU Psc b spectrum is very similar to that of J1021 (also shown on Figure \ref{fig:specstand}(b)), which was the initial standard for the T3 spectral type \citep{Burgasser2002}, until it was confirmed through \textit{Hubble Space Telescope} NICMOS imaging to be a 0\farcs172\plmo0\farcs005 (5.0\plmo0.7)\,AU binary composed of a T1+T5 \citep{Burgasser2006b}. The GU Psc b spectrum is also very similar to J121440.95+631643.4 (J1214 hereinafter), a T3.5 discovered by \citet{Chiu2006}. \citet{Geissler2011} identified J1214 as a candidate binary because they obtained a better fit with a composite spectrum made of T2 and T6 templates than for a T3 template. Although GU Psc b remains significantly redder in $J$ -- \Ks, the $Y$ and $J$ bands are closely matched, and the fit is much better in $H$ band. In both cases, the only notable difference lies in the \Ks\-band spectrum. If GU Psc b is truly a single object, this suggests that it is slightly peculiar when compared to standards, perhaps due to the lower gravity expected for a relatively young object, or to some variation in metallicity or cloud properties.

\subsubsection{Direct Comparison with Atmosphere Models}
\label{subsec:directcomp}
To further constrain the physical properties of GU Psc b, we compared its NIR spectrum and its $z$, $W1$, and $W2$ photometry to the synthetic SEDs of two different sets of brown dwarf atmosphere models: 
the BT-Settl CIFIST model presented in \citet{Allard2013}\footnote{This grid, that uses the \citet{Caffau2011} solar abundance, is available online at \\ \url{phoenix.ens-lyon.fr/Grids/BT-Settl/CIFIST2011/}.} and the model presented in \citet{Morley2012}, which include low-temperature condensates (primarily sulfides). To determine quantitatively the best fit to this mix of photometric and spectroscopic data points, we used a method similar to that presented in \citet{Cushing2008} in order to compute a goodness-of-fit $G_{k}$ for each model $k$ that is minimized for the best fitting models :

\begin{equation}
G_{k}=\sum_{i=1}^{n}W_i\left (\frac{F_{\rm{obs},i}-C_{k}F_{k,i}}{\sigma_{\rm{obs},i}}\right)^{2}.
\end{equation}

In this equation, $F_{\rm{obs},i}$ is the flux observed in a given spectral range $i$. The associated uncertainty, $\sigma_{\rm{obs},i}$, is dominated by the \env5\% photometric zero-point uncertainty. The total number of spectral ranges is $n$: one per photometric filter or spectral point (after smoothing).  $F_{k,i}$ is the flux of the synthetic model over the same wavelength domain. The synthetic model spectra are convolved at the resolution of the spectrum and the synthetic photometric magnitudes are obtained using instrumental filter profiles\footnote{The GMOS-South $z$ transmission curve available at \\ \url{www.gemini.edu/sciops/instruments/gmos/filters/}\\ \url{gmos_s_z_G0328.txt} is convolved with the detector response \\(\url{www.gemini.edu/sciops/instruments/gmos/imaging/} \\ \url{detector-array/gmoss-array}). The \textit{WISE} transmission curves are taken from \citet{Wright2010}.}. Rather than using $W_{i}$ = $\Delta\lambda$ = $\lambda_{2}-\lambda_{1}$ for the weight (as done in \citealt{Cushing2008}),  we chose to use $W_{i}$= $\Delta\log\lambda$ = $\log(\lambda_{2}/\lambda_{1})$  to purposefully \textit{not} be biased by the arbitrary choice of working in wavelength space rather than frequency space. The scale $C_{k}$ is equal to the dilution factor $R^{2}/d^{2}$ for a source of radius $R$ at a distance $d$. 
For each model, we constrained $C_{k}$ using the central distance inferred from Bayesian statistical analysis for GU Psc in ABDMG, \ds\ = 48\,pc, and the radius prescribed at the given \Teff\ and \logg\ by evolution models (\citealt{Saumon2008} for the low-temperature cloud model and \citealt{Baraffe2003}\footnote{Available at \url{phoenix.ens-lyon.fr/Grids/BT-Settl/} \\ \url{CIFIST2011/ISOCHRONES/}.} for BT-Settl). 

\begin{figure}[hbtp]
\begin{center}
\includegraphics[width=8cm]{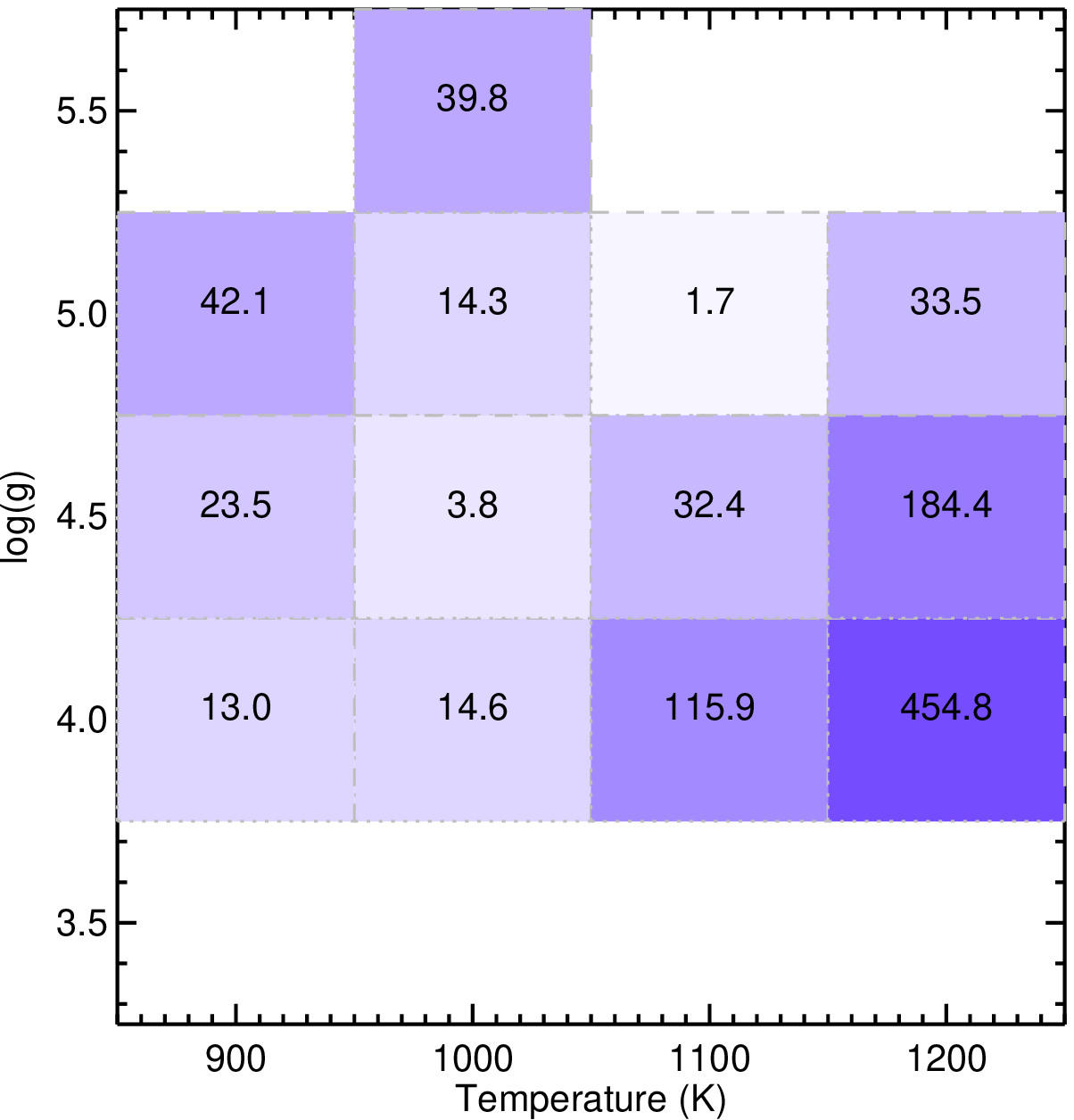}

\includegraphics[width=8cm]{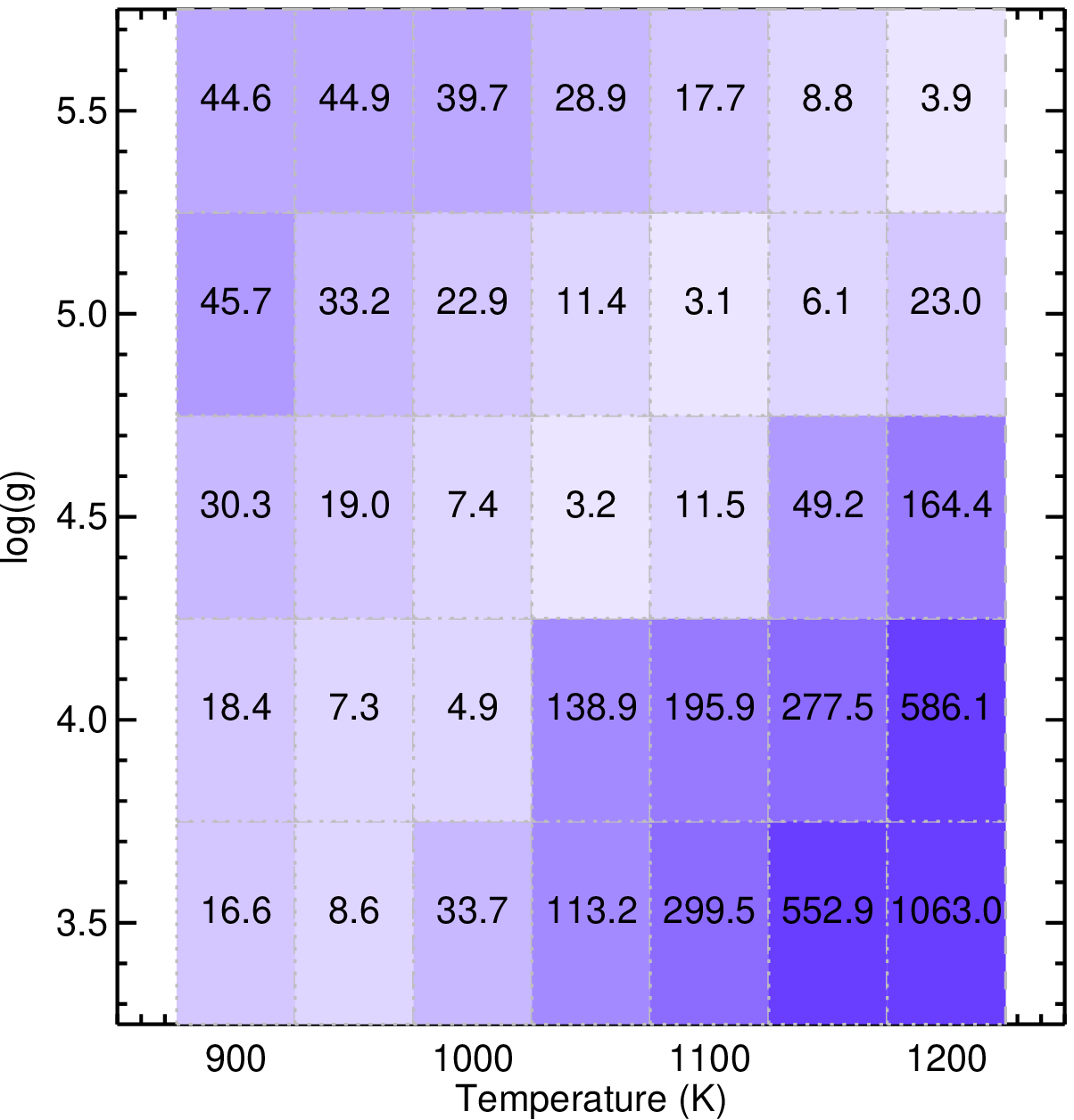}
\caption{Goodness-of-fit maps for the low-temperature cloud model of \citet{Morley2012} (upper panel) with \fsed\ = 1 and \Kzz\ = 0\,\cmdsmun and the BT-Settl model (lower panel), using the same contour levels. The value of the goodness-of-fit $G$ is written for each set of parameters. Both sets of models reach a minimum around \Teff\ = 1000--1100\,K and \logg\ = 4.5--5.0. These results are obtained with the scale $C_{k}$ constrained using the radius from evolutionary models (\citealt{Saumon2008} for the low-temperature cloud model and \citealt{Allard2013} for the BT-Settl) and \ds\ = 48\,pc for the distance.}
\label{fig:specchi2map}
\end{center}
\end{figure}

For the low-temperature cloud model described in \citet{Morley2012}, we used a grid with temperatures between 700\,K and 1300\,K ($\Delta$\Teff\ = 100\,K) and \logg\ between 4.0 and 5.5 ($\Delta$\logg\ = 0.5), at solar metallicity. We also tried different values for the sedimentation efficiency \fsed\ (1--5) and for \Kzz, which quantifies departure from chemical equilibrium (0 and $10^{4}$\,\cmdsmun). For the BT-Settl model, we used the CIFIST grid presented in \citet{Allard2013}, computed with temperatures between 700\,K and 1400\,K ($\Delta$\Teff\ = 50\,K) and \logg\ between 3.5 and 5.5 ($\Delta$\logg \ = 0.5), also at solar metallicity. 

\begin{figure*}[htbp]
\begin{center}
\includegraphics[width=12cm]{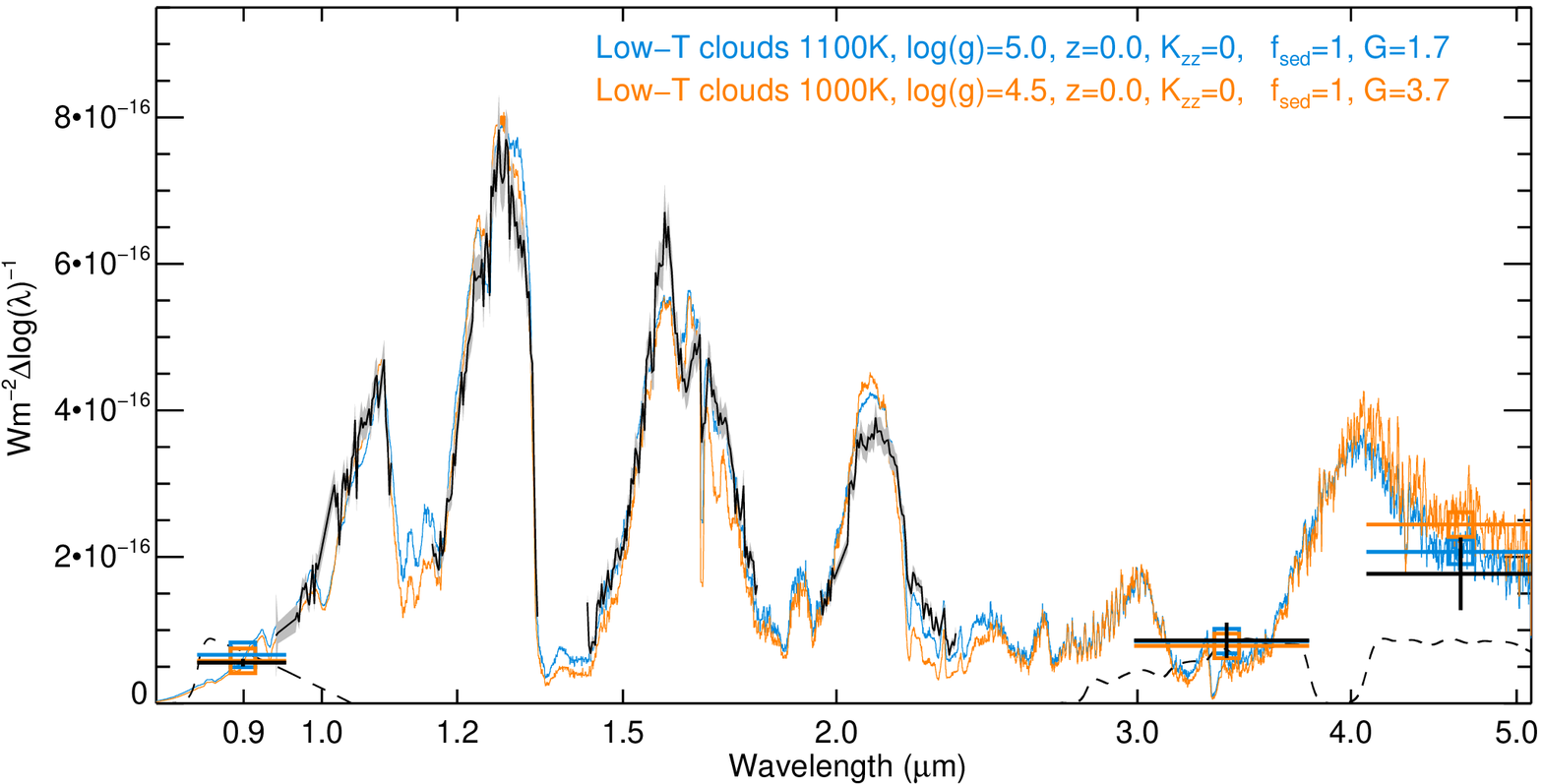}\\
\includegraphics[width=12cm]{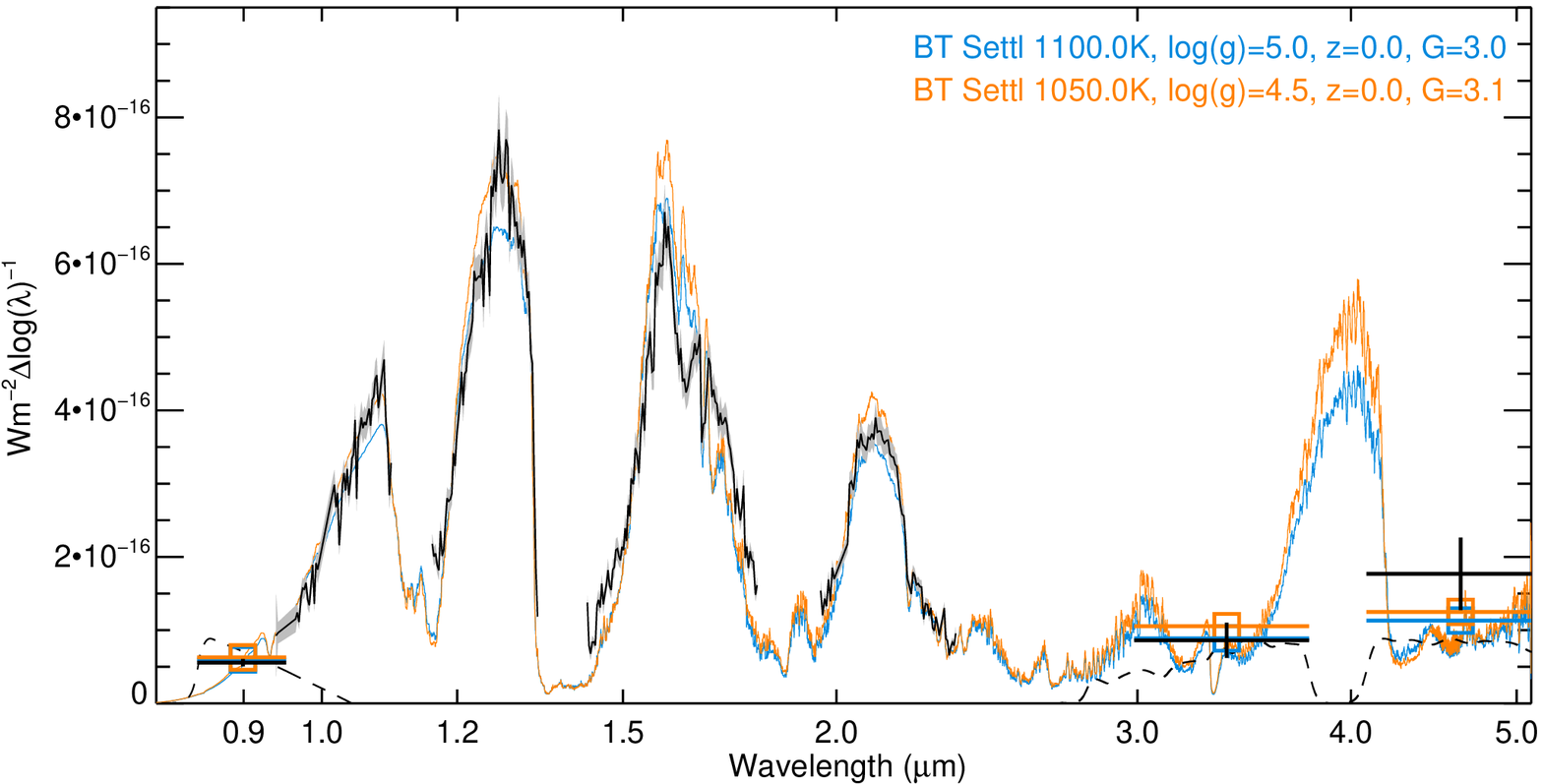}\\
\caption{GU Psc b GNIRS NIR spectrum and $z$, $W1$, and $W2$ photometry points along with the best-fit model spectra for the low-temperature cloud model of \citet{Morley2012} (upper panel) and the BT-Settl model (lower panel). For each model, the parameters (effective temperature \Teff, surface gravity \logg, metallicity $z$, the Eddy coefficient \Kzz\ and the sedimentation parameters \fsed) are given. The goodness-of-fit $G$, that allows us to quantify the quality of the fit (the smaller $G$ is, the better the fit; see Section \ref{subsec:directcomp}) is also shown. The model flux scale is absolute, using the radius from evolutionary models (\citealt{Saumon2008} for the low-temperature cloud model, and \citealt{Allard2013} for the BT-Settl) for every set of parameters, and the statistical distance of the primary in ABDMG, \ds\ = 48\,pc.}
\label{fig:specmodelab}
\end{center}
\end{figure*}

Figure \ref{fig:specchi2map} shows the goodness-of-fit map for both sets of models in the temperature/\logg\ parameter space. The physical parameters that lead to the best fit between the observed spectrum and the models are the same for both sets, \Teff\ \env\ 1000--1100\,K and \logg\ \env\ 4.5--5.0. The observed SED of the object constrains the bolometric luminosity, there is thus a correlation between temperature and gravity for the best fits: they are achieved either at a low temperature (\Teff\ = 1000--1050\,K) and low surface gravity (\logg\ = 4.5) or at higher temperature (\Teff\ = 1100\,K) and greater gravity (\logg\ = 5.0). In the case of BT-Settl, even higher temperature (\Teff\ = 1200\,K) and surface gravity (\logg\ = 5.5) still give a good fit, although these physical parameters are not consistent with the age of ABDMG (see Section \ref{sec:evolmod}). Figure \ref{fig:specmodelab} shows the GU Psc b SED and the two best synthetic spectra for both sets of models. 

We also did the entire fitting process constraining the $C_{k}$ with the extreme values of the distance range, 43\,pc and 53\,pc. The effective temperature and surface gravity we obtained did not vary significantly.

\begin{figure*}[hbtp]
\begin{center}
\includegraphics[width=12cm]{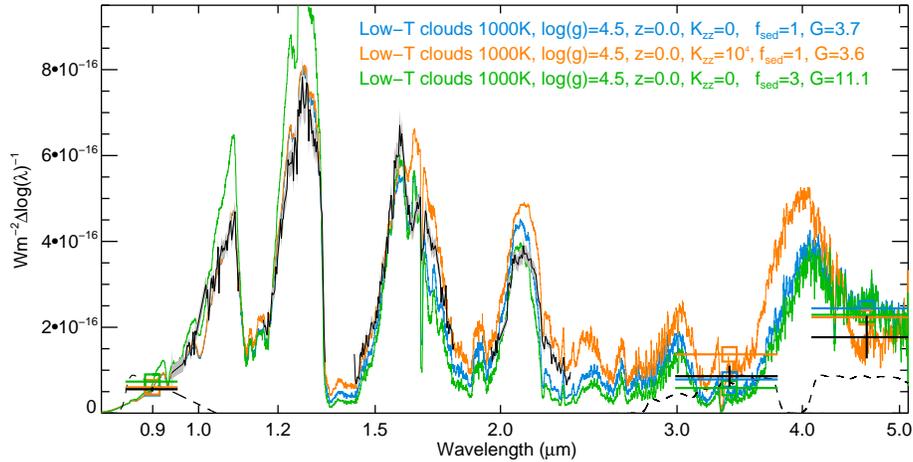}
\caption{Same as Figure \ref{fig:specmodelab}, only now the GNIRS spectrum and photometry points are compared to the low-temperature cloud model for various values of the parameters \fsed\ and \Kzz. }
\label{fig:specmodelc}
\end{center}
\end{figure*}

Both synthetic models match the SED of GU Psc b between 0.9 and 5\,\micron\ remarkably well, especially considering that there is no adjustment of the absolute flux of the models ($C_{k}$ is constrained). Overall, BT-Settl fits the $K$ band (especially its red side) and the $Y$ band around 1\,\micron\ slightly better than the low-temperature cloud model, but the later one better reproduces the $J-H$ color. The methane band at 1.6\,\micron\ is typically poorly matched by the models which tend to overestimate the flux in this wavelength range (see e.g., HN Peg b SED on Figure 4 of \citealt{Leggett2008}). Here, we observe an opposite trend: the flux redward of 1.7\,\micron\ is too low in both sets of models compared to GU Psc b. This excess flux, which is also seen in the comparison of GU Psc b to the T3 and T4 templates on Figure \ref{fig:specstand}(a), could be indicative of a slight departure from solar metallicity or an inhomogeneous surface, or an indication that GU Psc b is a very tight, unresolved, binary object.

The good fit we obtain with \citet{Morley2012} model is somewhat surprising, considering our object is hotter than the targeted objects for this model. This model came as an attempt to explain why late T dwarfs (\Teff\lta900\,K, primarily) were not perfectly fitted with models without iron and silicate clouds. Using \citet{Ackerman2001} cloud model, they studied the influence of other condensates (sulfides mainly, e.g., Na$_{2}$S, MnS, and ZnS, but also KCl and Cr), that were previously ignored and realized they are important at low temperatures (between 400 and 1300\,K). 
Since the targeted objects are late T dwarfs, \citet{Morley2012} models do not include the iron and silicate clouds that are important for L dwarfs since they are thought to be rapidly clearing at the L/T transition. In models including such condensates (but not the ones included in the \citealt{Morley2012} model), early T such as GU Psc b are expected to be best described by thin clouds of larger particles (corresponding to a high \fsed\ parameter in the \citealt{Ackerman2001} model). This is the case, for example, for HN Peg b (T2.5),  for which the best fit is obtained with \fsed\ = 3.5 \citep{Leggett2008} or for J1254 (T2) or 2M J05591914--1404488 (T4.5), which were best fit with \fsed\ = 3 and 4, respectively, in \citet{Cushing2008}. 
Alternatively, our best fits with \citet{Morley2012} model are obtained using \fsed\ = 1, thus very thick clouds of the less abundant sulfide condensates provide the moderate dust opacity evident in our T3.5 object. Figure \ref{fig:specmodelc} compares one of our best fits using \fsed\ = 1 (in blue) and a much poorer fit with the same temperature/\logg\ but with thinner sulfide clouds (\fsed\ = 3, in green). 

The second parameter is the Eddy diffusion coefficient, \Kzz\ that characterizes the vertical transport in the atmosphere. Vigorous vertical transport can bring molecular species from deeper, hotter layers of the atmosphere to the upper, cooler layers on a time scale faster than that of some chemical reactions, driving the molecular abundances away from their local equilibrium values. In particular, this results in increased CO and $\rm{CO}_2$ abundances, and reduced \CHq, \HdO, and \NHt\ abundances in the upper atmosphere \citep{Lodders2002, Saumon2006, Burningham2011}. Figure \ref{fig:specmodelc} shows, for \fsed\ = 1, the two available values we tested for the Eddy diffusion coefficient: \Kzz\ = 0\,\cmdsmun\ (chemical equilibrium, in blue) and \Kzz\ = $10^{4}$\,\cmdsmun\ (in orange).  It shows that \Kzz\ has little impact on the $Y$ and $J$ bands, but that an increase in \Kzz\ increases the flux in $H$, $K$ and in the mid-infrared. The best fit at \Kzz\ = 0\,\cmdsmun\ is obtained at \Teff\ = 1100\,K and \logg\ = 5.0 ($G$ = 1.7, Figure \ref{fig:specmodelab}, in blue) while at \Kzz\ = $10^{4}$\,\cmdsmun, the best fit is obtained at a slightly lower temperature and \logg: \Teff\ = 1000\,K and \logg\ = 4.5 ($G$ = 3.6, Figure \ref{fig:specmodelc}, in orange). The depth of the 4.6\,\micron\ absorption band in $W2$ is reproduced a bit better with a higher values of \Kzz. We must caution that we did not try a higher \Kzz\ for values of \fsed\ different than 1.

The BT-Settl model treats condensation and sedimentation of all dust species as well as gas phase advection by relating to a single atmospheric velocity field derived from radiation hydrodynamical models. There are thus no adjustable cloud parameters in this model. The cloud sedimentation and Eddy diffusion are instead determined by direct comparison of the relevant timescales (for condensation, sedimentation, chemical reactions) to the mixing timescale derived from this velocity field. The cloud opacity is composed of a number of condensates, which are settling to various degrees. Although the current version of the model does not yet include the opacity contribution of all low-temperature condensates that are included in \citet{Morley2012} (most notably Na$_{2}$S), the BT-Settl model reproduces the observed spectrum of GU Psc b similarly well. This is because even if these low-temperature clouds start to form at high altitude for L/T transition objects like GU Psc b, they do not become optically thick in this effective temperature range. The continuum of the flux peaks is still shaped by the silicate clouds, even if they have receded relatively deep into the photosphere.

The excess absorption at 4--5\,\micron\ apparent for this model could be indicative of an overestimation of the diffusion efficiency, resulting in too much CO and CO$_{2}$ being mixed into the photosphere. The mixing derived at the transition from CO- to CH$_4$-dominated atmosphere regions yields a diffusion coefficient of \env10$^5$--10$^6$\,\cmdsmun. It thus produces more CO and CO$_2$ absorption at 4--5\,\micron\ than the \Kzz\ = 0 or the \Kzz\ = 10$^4$ models of \citet{Morley2012}. Alternatively, the excess flux in the CH$_4$ absorption band in $W1$ might also reveal the incompleteness of the currently used CH$_4$ opacities, which cover only a fraction of the lines relevant at temperatures above 1000\,K \citep{Yurchenko2014}.

We investigated the effect of leaving the scale $C_{k}$ free in the fitting. For both synthetic models, for a given set of parameters, we then obtain a similar or slightly better fit than when imposing the scale. Nonetheless, the best fit occurs for the same physical parameters than with a constrained $C_{k}$. This strengthens our confidence both in the radii predicted by evolutionary models and in the distance inferred by the Bayesian statistical analysis.

\subsubsection{Physical Properties from Evolutionary Models}
\label{sec:evolmod}
Evolutionary models can be used to constrain the physical parameters (radii, surface gravity, bolometric luminosity and mass) of GU Psc b. In the previous section, we showed that the best fit were obtained with \Teff/\logg\ couples of 1000--1050\,K/4.5 or 1100\,K/5.0. Evolutionary models suggest that the lowest temperature and \logg\ are the most likely; the age is then consistent with the age of ABDMG, 100\plmo30\,Myr. The highest temperature and \logg\ would imply ages older than \env300\,Myr. The age deduced from evolutionary models for 1200\,K/5.5, which also led a good fit for BT-Settl, are greater than 1\,Gyr and are clearly excluded for ABDMG age.  

Thus, using ABDMG age (100\plmo30\,Myr) and the range of plausible temperatures determined with atmosphere models (\Teff\ = 1000--1100 K) in evolutionary models, we obtain the ranges of values for the physical properties of GU Psc b presented in Table \ref{table:derivedprop}. We used the two different models presented in \citet{Saumon2008} : one with \fsed\ = 2, which is a good approximation for all cloudy models, and one without clouds. We present both results as limiting cases in Table \ref{table:derivedprop}, but since the atmosphere model fitting suggests a better match with clouds than without, the cloudy version is probably the most appropriate for GU Psc b. We also used the evolutionary model of \citet{Baraffe2003} and obtained similar results. 
In all cases, the values obtained for the bolometric luminosity are between \Lbol\ = --4.9 and --4.6. The  surface gravity inferred  (\logg\ = 4.2--4.4) is consistent, albeit slightly lower, with the values derived from atmosphere model fitting (\logg\ = 4.5--5). All models suggest a mass between 9 and 13\,\MJ. We thus find that GU Psc b is probably below the lower threshold of deuterium burning for its \Teff\ and age, unlike 2M0122 and several other brown dwarfs discussed in \citet{Bowler2013} that are possible deuterium burners.

\begin{table}
\begin{center}
 \newcolumntype{A}[1]{>{\arraybackslash}m{#1}} 
 \newcolumntype{W}[1]{>{\centering\arraybackslash}m{#1}}  
\caption{Evolutionary Model Derived Physical Properties of GU Psc b\tnma}
\label{table:derivedprop}
\begin{tabular}{W{1.9cm}@{\hskip 0.1cm}W{1.5cm}@{\hskip 0.2cm}W{1.5cm}@{\hskip 0.2cm}W{1.7cm}@{\hskip 0.2cm}W{1.5cm}} 
\hline \hline   
&&&&\\
 Property    & $R$       &\logg\ & \Lbol\  & Mass \\
    & (\RJ)    &                &            &   (\MJ)   \\
 Model     &&&&\\
\hline 
&&&&\\
S08 cloudy\tnmb       & 1.33--1.38   				& 4.18--4.23 				& --4.80---4.60 		& 10.8--12.0\\
S08 no cloud\tnmb   & 1.23--1.27   				& 4.23--4.31 				& --4.87---4.67 		& 10.4--12.1\\
B03\tnmc   		  &1.15--1.21   				&  4.22--4.36 			& --4.91---4.70 		& 9.6--13.0\\
\hline
\multicolumn{5}{A{8.1cm}} {}\\ 
\multicolumn{5}{A{8.1cm}} {\textbf{Notes}}\\ 
\end{tabular}
\begin{tabular}{p{0.1cm}p{8.0cm}}
\tnma & Assuming an age of 100\plmo30\,Myr and a \Teff\ = 1000--1100 K.\\ 
\tnmb & Using \citet{Saumon2008} evolutionary models. The \textit{cloudy} version is with \fsed\ = 2, and is also appropriate for \fsed\ = 1 (see text). \\
\tnmc & Using \citet{Baraffe2003} evolutionary models.\\
\end{tabular}
\end{center}
\end{table}

\section{Analysis and Discussion}
\label{sec:analdis}

\subsection{Stability of the System}
This system has a very wide projected separation ($r$ = 2000\plmo200\,AU) that is not seen in any other planet--star system (excluding companions to white dwarf or other evolved systems such as WD 0806--661, \citealt{Luhman2011};  LSPM1459+0857, \citealt{DayJones2011} or sdM1.5+WD Wolf 1130, \citealt{Mace2013} for which stellar mass loss most likely had an impact on the separation). The mass ratio of GU Psc system ($q\sim0.03$) is higher than that of typical exoplanetary systems. The value is particularly high for an M dwarf host, these stars seem to be an uncommon host for Jupiter-mass companions, even at close separations \citep{Bonfils2011}. However, it is significantly lower than the mass ratio of several directly imaged systems, such as the 30\,\MJ\ brown dwarf around the M1 star CD-35 2722 ($q\sim0.07$; \citealt{Wahhaj2011}) or the 4\,\MJ\ around the M8 brown dwarf 2M 1207 ($q\sim0.15$; \citealt{Chauvin2004}). 

The very large separation coupled to the very low mass ratio lead to a very small binding energy. With a primary mass of $M_{\star}$ = 0.30--0.35\,\Msun, using $M_{c}$ = 9--13\,\MJ\ for the mass of the  companion and $r$ = 2000\,AU as the instantaneous projected separation, the binding energy of the system is (assuming a circular orbit)

\begin{equation}
E_{\rm{bind}}\sim -\frac{GM_{\star}M_c}{1.27r}=-(0.2\pm0.1)\times 10^{41}\,\rm{erg},
\end{equation} 

where 1.27 is the average projection factor between $r$ and the semimajor axis, assuming a random viewing angle, see, e.g., \citet{Brandeker2006}. Although this binding energy is very small, it is the same order of magnitude as that of other presumably gravitationally bound systems that include a planetary-mass companion, such as Ross 458(AB) c system (\Ebind\ \env\ -1\ $\times$ $10^{41}$\,erg, using $r$ \env\ 1168\,AU,  $M_{\star}$ \env\ 0.61\,\Msun, and $M_{c}$ \env\ 14\MJ; \citealt{Goldman2010}) or the 2M1207 A brown dwarf (BD) and its companion (\Ebind\ \env -0.2--0.6  $\times$ $10^{41}$\,erg, using $r$ \env\ 52\,AU,  $M_{\rm{BD}}$ \env\ 25\,\MJ, and $M_{c}$ = 3--10\,\MJ; \citealt{Chauvin2004,Ducourant2008}). It is also similar to the binding energy of older, more massive systems such as the M3 star G204-39 and T6.5 brown dwarf 2MASS J1758+4633 (\Ebind\ \env --0.4--0.7 $\times$ $10^{41}$\,erg, using the masses and separation given in \citealt{Faherty2010}) and greater than many of the very-low mass star systems presented in \citet{Dhital2010}. Thus, it is not unreasonable for GU Psc b to be gravitationally bound to its primary despite its very wide separation.
 
Indeed, one can estimate the average survival time of the system considering the encounters with stars and giant molecular clouds, which are the most important sources of disruption. The chance of disruption depends mainly on the binding energy, so the results shown in Figure 2(a) of \citet{Weinberg1987} for a 1\,\Msun\ system can be scaled down for GU Psc system. With a projected separation of $r$ \env\ 2000\,AU (\env9.7 $\times$ $10^{-3}$\,pc) and a total mass of \env\ 0.35\,\Msun, our system lies between the $a_{0}$ = 0.02\,pc and $a_{0}$ = 0.04\,pc curves, which implies a half-life for the system of \env\ 5--6\,Gyr, which is much greater than its estimated age.

\subsection{Formation Mechanisms} 
With such a great distance from its host star and a relatively high mass ratio, it is unlikely that GU Psc b was formed alone in situ in a protoplanetary disk, through one of the canonical formation mechanisms for exoplanets (core accretion; \citealt{Pollack1996,Inaba2003} or disk instability; \citealt{Bate2002,Boss2006,Stamatellos2007,Rafikov2009}). It would require too large a protoplanetary disk, with an unrealistically large density at this separation. 

A plausible scenario is that GU Psc b formed in a disk, but migrated outward due to dynamic interactions with an unseen, more massive companion to GU Psc A \citep{ReipurthClarke2001}. Similar low-mass, wide companions have been found around binary star (e.g., Ross 458(AB) c, at $>$ 1100\,AU, SR 12 AB c, at $>$ 1000\,AU).
This hypothesis would imply that GU Psc A has a more massive, closer-in companion that was not seen in our observations (see Figure \ref{fig:multiplicity}). More high-contrast imaging observations would be useful to better constrain the binary nature of GU PSc A. It would also be desirable to assess theoretically whether the hypothetical triple systems allowed in Figure \ref{fig:multiplicity} represent realistic dynamical stable solutions, given the age of the system. 

Another possibility is that GU Psc A and b were both formed in the disk of a more massive star, and have been ejected as a system \citep{Bate2003,Stamatellos2009}.

Both components of the system could also have been formed by turbulent core fragmentation, as a weakly bound binary system, similar to a binary star system. 
It would imply that cores can fragment into objects as light as the companion (which is plausible, given the opacity limit for fragmentation is a few Jupiter masses; \citealt{Bate2009}) or that the system was ejected from the accretion reservoir following a dynamical interaction \citep{ReipurthClarke2001, Bate2002, BateBonnell2005}. 

GU Psc b could also have been a free-floating planet, formed by turbulent core fragmentation or ejection from a protoplanetary system, later captured by GU Psc A \citep{Perets2012}. In this case, we would observe less correlations between the physical properties (e.g., metallicities or spin--orbit relation) of the primary and secondary. More insight on the possible formation mechanism could thus possibly be obtained by determining the relative metallicity of both components through high-resolution spectroscopy.

\subsection{Interest of the System}

\begin{figure*}[htbp]
\begin{center}
\includegraphics[width=14cm]{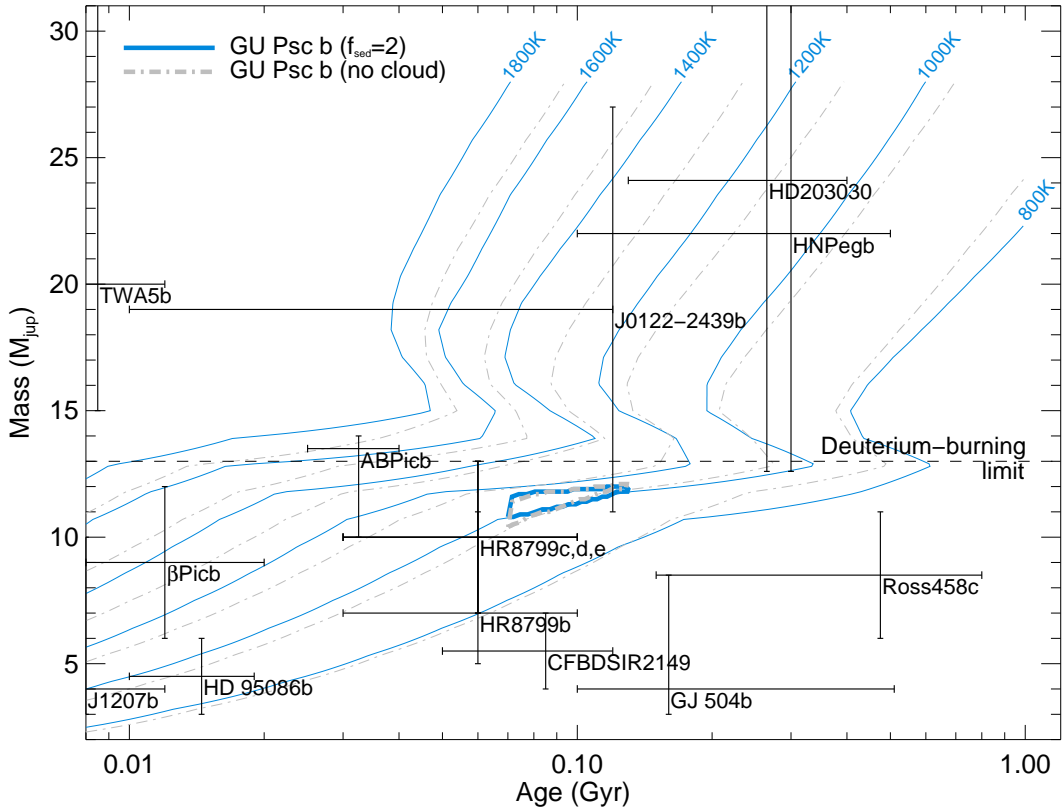}
\caption{Masses and ages of various low-mass companions. Evolutionary models of \citet{Saumon2008} with and without clouds are represented by the solid blue line and the dot-dashed gray line, respectively. Given the age of the ABDMG and the effective temperature range found using atmosphere models (\Teff\ = 1000--1100 K), the mass of GU Psc b is below 13\,\MJ, irrespective of the model used (10--12\,\MJ\ for \citealt{Saumon2008} model). Ages and masses for other sources are taken from \citet{Bonnefoy2010}, \citet{Bowler2013}, \citet{Burningham2011}, \citet{Chauvin2004}, \citet{Ducourant2008}, \citet{Kuzuhara2013}, \citet{Lagrange2010}, \citet{Liu2011}, \citet{Lowrance1999}, \citet{Luhman2007}, \citet{Marois2008}, \citet{Marois2010}, \citet{Metchev2006}, \citet{Neuhauser2012}, \citet{Rameau2013}, and \citet{Delorme2012}.}
\label{fig:LMOmassage}
\end{center}
\end{figure*}

Figure \ref{fig:LMOmassage} compares the masses, ages, and temperatures of GU Psc b and other low-mass companions. The temperature derived from atmosphere models for GU Psc b and the age of ABDMG yield a mass below the deuterium-burning mass limit in evolutive models(the \citealt{Saumon2008} model is shown on Figure \ref{fig:LMOmassage}). 

Figure \ref{fig:LMOmassage} also shows that GU Psc b is intermediate in age between the few planetary-mass objects uncovered in very young associations (e.g., the 8--20\,Myr-old planet 2M1207 b in TWA or the 12\,Myr-old $\beta$ Pictoris b) and the field planetary-mass objects of hundreds Myr and more (e.g., the cool 150--800\,Myr Ross 458(AB) c). While the constraints on the age of many companions come from the poorly known age of the system, the ABDMG membership of GU Psc provides much better constraints to the age of the companion, allowing a better validation of the models, still poorly constrained for this mass and age range. 

The similarity in age and mass between GU Psc b and the HR 8799 planets is obvious on Figure \ref{fig:LMOmassage}. Indeed, the physical properties of GU Psc b ought to be very similar to the most massive planets that should be discovered by forthcoming planet finder instruments such as the Gemini Planet Imager \citep{Macintosh2012}, SPHERE on the VLT \citep{Beuzit2008}, or the HiCIAO on Subaru \citep{Suzuki2009}, albeit at much closer separations. GU Psc b could serve as a proxy for these planets, that will likely be only characterizable with low-resolution spectroscopy ($R$ \env\ 40), due to their proximity to their parent star. GU Psc b, at \env\ 42\arcsec\ from its primary, will be amenable to detailed follow-up photometric and spectroscopic observations, just like the free-floating planetary-mass object CFBDSIR2149, also a strong candidate member of ABDMG (4--7\,\MJ; \citealt{Delorme2012}). 

GU Psc b follows the trend of companions like 2M1207b, 2M 0122B, the planets of HR 8799 and HN Peg B that all show effective temperatures under that of typical field brown dwarfs for a given spectral type  (see \citealt{Bowler2013}, Figure 13). This is likely explained by the lower surface gravity of these objects. GU Psc b provides an important data point to clarify the spectral type versus temperature relationship at intermediate ages.

With its spectral type of T3.5\plmo1, a derived temperature between 1000 and 1100\,K and an age of 70--130\,Myr, GU Psc b is a rare example of a {\it young} early T dwarfs straddling the L/T transition between the cloudy and clear regime of brown dwarf atmospheres. The L/T transition is particularly challenging for atmosphere models because of the complex treatment of clouds required. At this transition, it is expected that the iron/silicate clouds, important source of opacity for L dwarfs \citep{Saumon2008,Stephens2009}, gradually become less important, either because they sink or become patchy, which allows the emergent flux to come from a deeper layer of the atmosphere. As suggested in \citet{Morley2012}, other condensate could become important for cooler mid- to late-T dwarfs. As shown in Section \ref{subsec:directcomp}, GU Psc b's SED is well reproduced  both by the BT-Settl model \citep{Allard2012, Allard2013}, that includes iron/silicate condensates (and some of the low-temperature condensates), and by the \citet{Morley2012} model that includes sulfides and other low-temperature condensates, using thick clouds (\fsed\ = 1). 

Several early field T dwarfs are known to show a photometric variability that could, in fact, be explained by a combination of cloudy and clear regions in the atmosphere, or by a partial cloud cover.
For GU Psc b, our limited set of three $J$-band images (spanning \env11 months; see Table\,\ref{table:prop}) gives magnitudes that are all consistent with each other within 1$\sigma$. These data set a 3$\sigma$ upper limit on variability of approximately 150\,mmag. GU Psc b would be a prime target for further photometric variability studies. Although it is much fainter than other field early T dwarfs, variability studies of GU Psc b are well within the capability of existing ground- and space-based telescopes.

The binary hypothesis being to a large extent ruled-out for the companion, it would be interesting to extend the analysis with models atmospheres and spectra for partly cloudy atmospheres \citep{Marley2010} or even composite spectra for atmosphere models with horizontal temperature variations.

\section{Summary and Conclusions}
\label{sec:sumcon}
We have presented the discovery of a co-moving planetary-mass companion to GU Psc, a low-mass M3 star and strong candidate member of the \env100\,Myr old ABDMG association. We presented evidences that strongly support a membership in ABDMG. Notably, its kinematics and X-ray emission fit that of the association. The companion is widely separated from the host star at \env42\arcsec\ or \env2000\,AU at the estimated distance of 48\,pc. 

The companion has the spectral signature of a T3.5\plmo1 spectrum, with relatively strong $K$-band emission, a likely indicator of a low-gravity object. The overall SED resembles closely that of a  binary T0+T5 (J1021) and of a candidate binary T2+T6 (J1214) spectra, but it has been shown with Keck LGS AO observations that it is in all likelihood a ``single'' early T3.5 dwarf. Few such L/T transition dwarfs are known. They constitute particularly interesting candidates for variability studies as they are likely to have partial and variable cloud cover. GU Psc b is a prime target to extend previous photometric variability studies of old early T dwarfs to younger ages.

Astrometric observations, through the CTIOPI\footnote{\url{www.chara.gsu.edu/~thenry/CTIOPI/}} project, are ongoing to secure the parallax of the system.
A precise distance will confirm the membership of the primary, hence confirm the age of the system, and allow to better establish the physical parameters of the companion. High-contrast imaging observations of the host star could put better constraints on the mass ratio and separation of an hypothetical closer companion to the star. 
The mid-infrared spectroscopy of GU Psc b could provide significant constraints to atmosphere models in this region. It will be an easy target for NIRSpec and MIRI on board the \textit{James Webb Space Telescope} \citep{Gardner2006}. GU Psc b should become an excellent proxy for relatively massive gas giant planets soon to be discovered by forthcoming high-contrast imaging instruments. 

\subsection*{Acknowledgments}
We would like to thank the anonymous referee for constructive comments and suggestions that greatly improved the overall quality of the paper.
This work was financially supported by the Natural Sciences and Engineering Research Council (NSERC) of Canada and the Fond de Recherche Qu\'{e}b\'{e}cois -- Nature et Technologie (FRQNT; Qu\'{e}bec). 
D.S is supported by NASA Astrophysics Theory grant NNH11AQ54I.
D.H. acknowledges support from the European Research Council under the European
Community's Seventh Framework Programme (FP7/2007-2013 Grant Agreement no. 247060).
Based on observations obtained at the Gemini Observatory (Gemini-S/PHOENIX: program GS-2010B-Q-89, Gemini-S/GMOS: program GS-2011B-Q-74, Gemini-S/NICI: programs GS-2011B-Q-24 and GS-2012B-Q-54 and Gemini-N/GNIRS: program GN-2012B-Q-58), which is operated by the Association of Universities for Research in Astronomy, Inc., under a cooperative agreement with the NSF on behalf of the Gemini partnership: the National Science Foundation (United States), the National Research Council (Canada), CONICYT (Chile), the Australian Research Council (Australia), Ministerio da Ciencia, Tecnologia e Inovacao (Brazil) and Ministerio de Ciencia, Tecnologia e Innovacion Productiva (Argentina). 
Observations were also collected at CFHT with WIRCam (run IDs: 11BC20 and 12BC20) and ESPaDOnS (run ID: 12AC23), 
at the European Southern Observatory Very Large Telescope under program ID: 087.D-0510, 091.D-0641 and on
CPAPIR infrared camera, at Observatoire du mont M\'{e}gantic, which is funded by the Universit\'{e} de Montr\'{e}al, Universit\'{e} Laval and the Canada Foundation for Innovation. 
Some of the data presented herein were obtained at the W. M. Keck Observatory, which is operated as a scientific partnership among the California Institute of Technology, the University of California, and the National Aeronautics and Space Administration. The Observatory was made possible by the generous financial support of the W. M. Keck Foundation. The authors recognize and acknowledge the very significant cultural role and reverence that the summit of Mauna Kea has always had within the indigenous Hawaiian community. We are most fortunate to have the opportunity to conduct observations from this mountain.
Finally, we also obtained data from the NASA InfraRed Telescope Facility, with SpeX, under the program number 2013B025.
This publication makes use of data products from the \textit{Wide-field Infrared Survey Explorer}, which is a joint project of the University of California, Los Angeles, and the Jet Propulsion Laboratory/California Institute of Technology, funded by the National Aeronautics and Space Administration, from the Two Micron All Sky Survey, which is a joint project of the University of Massachusetts and the Infrared Processing and Analysis Center, and funded by the National Aeronautics and Space Administration and the National Science Foundation, of the NASA's Astrophysics Data System Bibliographic Services, SIMBAD database, the VizieR catalog access tool and the SIMBAD database operated at CDS, Strasbourg, France.
The BT-Settl model atmospheres have been computed at the P\^{o}le Scientifique de Mod\'{e}lisation Num\'{e}rique of the ENS de Lyon, and at the Gesellschaft f\"{u}r Wissenschaftliche Datenverarbeitung G\"{o}ttingen in co-operation with the Institut f\"{u}r Astrophysik G\"{o}ttingen.
This publication has made use of the L and T dwarf data archive, \url{staff.gemini.edu/~sleggett/LTdata.html}.



\end{document}